\documentclass[fleqn,usenatbib]{mnras}

\usepackage{newtxtext,newtxmath}

\usepackage[T1]{fontenc}

\DeclareRobustCommand{\VAN}[3]{#2}
\let\VANthebibliography\thebibliography
\def\thebibliography{\DeclareRobustCommand{\VAN}[3]{##3}\VANthebibliography}

\usepackage{lipsum}
\usepackage{subcaption}
\usepackage{multirow}

\usepackage{graphicx}	
\usepackage{amsmath}	
\graphicspath{{figures/}}

\usepackage[usenames,dvipsnames,svgnames,table]{xcolor}

\title[CMB lensing-galaxy cross correlation]{Cross-correlation between \textit{Planck} CMB lensing potential and galaxy catalogues from HELP}

\author[C. S. Saraf et al.]{
Chandra Shekhar Saraf,$^{1}$\thanks{E-mail: cssaraf@camk.edu.pl (CSS)}
P. Bielewicz,$^{2}$
M. Chodorowski$^{1}$
\\
$^{1}$Nicolaus Copernicus Astronomical Centre, Polish Academy of Sciences, ul.~Bartycka 18, Warsaw 00-716, Poland\\
$^{2}$National Centre for Nuclear Research, ul.~L.~Pasteura 7, Warsaw 02-093, Poland\\
}

\date{Accepted XXX. Received YYY; in original form ZZZ}

\pubyear{2021}

\begin{document}
\label{firstpage}
\pagerange{\pageref{firstpage}--\pageref{lastpage}}
\maketitle

\begin{abstract}
We present the study of cross-correlation between Cosmic Microwave Background (CMB) gravitational lensing potential map released by the \textit{Planck} collaboration and photometric redshift galaxy catalogues from the \textit{Herschel} Extragalactic Legacy Project (HELP), divided into four sky patches: NGP, \textit{Herschel} Stripe-82, and two halves of SGP field, covering in total $\sim 660$ deg$^{2}$ of the sky. We estimate the galaxy linear bias parameter, $b_{0}$, from joint analysis of cross-power spectrum and galaxy auto-power spectrum using Maximum Likelihood Estimation technique to obtain values ranging from $0.70 \pm 0.01$ for SGP Part-2 to $1.02 \pm 0.02$ for SGP Part-1 field. We also estimate the amplitude of cross-correlation and find the values spanning from $0.67 \pm 0.18$ for SGP Part-2 to $0.80 \pm 0.23$ for SGP Part-1 field, respectively. For NGP and SGP Part-1 fields the amplitude is consistent with the expected value for the standard cosmological model within $\sim 1\,\sigma$, while for \textit{Herschel} Stripe-82 and SGP Part-2 we find the amplitude to be smaller than expected with $\sim 1.5\,\sigma$ and $\sim 2\,\sigma$ deviation, respectively. We perform several tests on various systematic errors to study the reason for the deviation, however, value of the amplitude turns out to be robust with respect to these errors. The only significant change in the amplitude is observed when we replace the minimum-variance CMB lensing map, used in the baseline analysis, by the lensing map derived from the CMB temperature map with deprojected thermal Sunyaev-Zeldovich signal. 

\end{abstract}




\section{Introduction}

The Cosmic Microwave Background (CMB) has given us valuable insights into our Universe and the parameters that govern it and its evolution. We have entered an era of precision cosmology over the last two decades and one of the key contributors to this was the precise measurements of the CMB (\citeauthor{Planck2020I} \citeyear{Planck2020I}; \citeauthor{Planck2020VI} \citeyear{Planck2020VI}; \citeauthor{Planck2020VIII} \citeyear{Planck2020VIII}). This has helped to establish the standard model of cosmology and theory of large-scale structure (LSS) formation. Inhomogeneities generated during
the period of inflation developed into the structures we see today through gravitational collapse. Therefore, by studying the anisotropies in the CMB and large-scale structure,
we are able to infer the characteristics of the Universe at primordial times.

CMB photons traveling from the surface of the last scattering to us get deflected by matter inhomogeneities - an effect known as gravitational lensing. This effect changes the observed statistical properties of the CMB and alters the picture of the early Universe. On the other hand, these distortions carry also information about the LSS. The gravitational lensing causes deflections of CMB photons, with a typical amplitude of $2'$ \citep{Challinor2006}. These statistical signatures of the lensed CMB field can be exploited to reconstruct the lensing potential of the matter overdensities (\citeauthor{Zaldarriaga1999} \citeyear{Zaldarriaga1999}; \citeauthor{Hu2001} \citeyear{Hu2001}; \citeauthor{Hu2002} \citeyear{Hu2002}) and learn about the large-scale structure of the Universe \citep{Challinor2006}.

The map of the projected lensing potential can be reconstructed from CMB temperature and
polarisation data (\citealt{SPTPol2021};~\citealt{ACTPol2021};~\citealt{Planck2020VIII};~\citealt{Omori2017}). Since the CMB lensing is an integrated quantity along the line of sight, it does not provide direct information on the evolution of the large-scale gravitational potential. However, this information can be obtained from cross-correlation between the lensing map of CMB and tracers of LSS with known redshift. Since galaxies reside in dark matter halos \citep{Mo2010} they are good tracers of structures causing gravitational lensing of CMB. Cross-correlation studies can be used to determine the amplitude of structure at different redshifts (\citeauthor{Peacock2018} \citeyear{Peacock2018}; \citeauthor{Doux2018} \citeyear{Doux2018}), measure galaxy groups and cluster masses (\citeauthor{Gupta2021} \citeyear{Gupta2021}; \citeauthor{Raghunathan2019} \citeyear{Raghunathan2019}; \citeauthor{Planck2016XXIV} \citeyear{Planck2016XXIV}) and study the relation of luminous and dark matter (\citeauthor{Han2019} \citeyear{Han2019}; \citeauthor{Bianchini_Reichardt2018} \citeyear{Bianchini_Reichardt2018}).

Many cross-correlation studies have been performed over the past with optical catalogues like Sloan Digital Sky Survey (SDSS) \citep{Sukhdeep2020}, Dark Energy Survey (DES) (\citeauthor{Omori2019Cross} \citeyear{Omori2019Cross}; \citeauthor{Omori2019Tomo} \citeyear{Omori2019Tomo}), Wide-Field Infrared Survey Explorer (WISE) (\citeauthor{Krolewski2021} \citeyear{Krolewski2021}; \citeauthor{Krolewski2020} \citeyear{Krolewski2020}; \citeauthor{Goto2012} \citeyear{Goto2012}), Two Micron All Sky Survey (2MASS) \citep{Bianchini2018} and Subaru Hyper Suprime-Cam (for example, \citeauthor{Marques2020} \citeyear{Marques2020}; \citeauthor{Namikawa2019} \citeyear{Namikawa2019}) and with radio catalogues from Low-Frequency Array (LOFAR) \citep{Alonso2021}. Most of the wide-area galaxy surveys in visible/near-infrared or radio wavebands have objects with redshift slightly greater than one. Thus, only a fraction of the CMB lensing signal can be picked up through cross-correlations. Many such studies have been reported by a number of authors involving CMB lensing and galaxy density maps (\citeauthor{Darwish2021} \citeyear{Darwish2021}; \citealt{Cao2020};~\citeauthor{Aguilar2019} \citeyear{Aguilar2019}; \citeauthor{Giusarma2018} \citeyear{Giusarma2018}; \citeauthor{Schmittfull2018} \citeyear{Schmittfull2018}; \citeauthor{Pullen2015} \citeyear{Pullen2015}; \citeauthor{Giannantonio2015} \citeyear{Giannantonio2015}; \citeauthor{Kuntz2015} \citeyear{Kuntz2015}). Cross-correlation studies have also been reported between CMB lensing and quasar density maps (\citeauthor{Zhang2021} \citeyear{Zhang2021}; \citeauthor{DiPompeo2015} \citeyear{DiPompeo2015}; \citeauthor{Han2019} \citeyear{Han2019}) as well as high-redshift sub-millimeter sources from the \textit{Herschel} Astrophysical Terahertz Large Area Survey (H-ATLAS) have also been used to perform such studies (\citeauthor{Bianchini2015} \citeyear{Bianchini2015} and \citeauthor{Bianchini2016} \citeyear{Bianchini2016}).

In this paper, we present the first study of cross-correlation between \textit{Planck} CMB lensing potential \citep{Planck2020VIII} and galaxy catalogues from \textit{Herschel} Extragalactic Legacy Project (HELP; \citealt{Raphael2019,Raphael2021}). HELP catalogue is a combination of many surveys, mostly in visible and near-infrared bands, which also includes galaxies observed in \textit{Herschel} fields. As such it does not overlap with the H-ATLAS catalogue consisting mostly of far-infrared objects. The number of objects from the HELP catalogue in our study is orders of magnitude larger compared to the catalogue used by \cite{Bianchini2015} and \cite{Bianchini2016}.

We present the theoretical background in section \ref{sec:theory} and describe the \textit{Planck} lensing data and HELP data used in our study in section \ref{sec:data}. In section \ref{sec:methodology}, there are presented the procedure for estimation of power spectra and errors, as well as described our numerical setup employed for validation of the procedure. The method used for the estimation of parameters is described in section \ref{sec:likeli}. Finally we present our results in section \ref{sec:results} with further discussions in section \ref{sec:discuss}. At last, we summarise our results in section \ref{sec:summary}.

In this paper, we adopt the flat $\Lambda$CDM cosmology with best-fit \textit{Planck} + \textit{WP} + highL + lensing cosmological parameters, as described in \cite{Planck2020VI}. Here, \textit{WP} refers to \textit{WMAP} polarisation data at low multipoles, highL is the high resolution CMB data from Atacama Cosmology Telescope (ACT) and South Pole Telescope (SPT) and lensing refers to the inclusion of \textit{Planck} CMB lensing data in the parameter likelihood.


\section{Theory}\label{sec:theory}

Gravitational lensing of CMB photons can be expressed as a remapping of the unlensed temperature anisotropies $\Theta$($\hat{\textbf{n}}$) in the direction $\hat{\textbf{n}}$ \citep{Challinor2006}:

\begin{equation}
\begin{split}
	\tilde{\Theta}(\hat{\textbf{n}}) &=	\Theta(\hat{\textbf{n}}+\nabla\phi(\hat{\textbf{n}}))\\
	&= \Theta(\hat{\textbf{n}})+\nabla^{a}\phi(\hat{\textbf{n}})\nabla_{a}\Theta(\hat{\textbf{n}})+\mathcal{O}(\phi^{2})
\end{split}
	\label{eq:lensing_unlensing}
\end{equation}
where $\tilde{\Theta}(\hat{\textbf{n}})$ is the lensed temperature anisotropies and $\phi(\hat{\textbf{n}})$ is the CMB lensing potential defined as:

\begin{equation}
	\phi(\hat{\textbf{n}}) = -2\int_{0}^{\chi_{*}}d\chi \frac{\chi_{*}-\chi}{\chi_{*}\chi}\Psi(\chi\hat{\textbf{n}},z(\chi))
	\label{eq:lensing_potential}
\end{equation}
In above equation, $\chi_{*}$ is comoving distance to the surface of last scattering at redshift $z\simeq 1100$ and $\Psi(\chi\hat{\textbf{n}},z(\chi))$ is the three dimensional gravitational potential at position $\chi\hat{\textbf{n}}$ in photon's path. The deflection angle is then given by the two dimensional gradient on the sphere, $\nabla_{\hat{\textbf{n}}}\phi$.

The effects of gravitational lensing due to foreground matter introduces small coherent distortions in the light coming from background sources. We define dimensionless lensing convergence $\kappa$ through relation to the two-dimensional Laplacian of the lensing potential:

\begin{equation}
    \kappa(\hat{\textbf{n}}) = -\frac{1}{2}\nabla^{2}\phi(\hat{\textbf{n}})
    \label{eq:2d_laplacian_relation}
\end{equation}
The lensing convergence in a given direction of the sky can be related to the line-of-sight average of the matter over-density $\delta$ \citep{Bartelmann2001}:

\begin{equation}
	\kappa(\hat{\textbf{n}}) =\int_{0}^{\chi_{*}}d\chi\frac{H(\chi)}{c}W^{\kappa}\delta(\chi\hat{\textbf{n}})
	\label{eq:convergence_delta_relation}	
\end{equation}
where $W^{\kappa}$ is the lensing kernel given by
\begin{equation}
	W^{\kappa}(\chi) = \frac{3\Omega_{m}}{2c^{2}}H_{0}^{2}(1+z)\chi\frac{\chi_{*}-\chi}{\chi_{*}}
	\label{eq:lensing_kernel}
\end{equation}
where $c$ is the speed of light, $\Omega_{m}$ and $H_{0}$ are the present-day values of the matter density parameter and Hubble constant, respectively.

Similarly, the galaxy over-density $g(\hat{\textbf{n}})$ can also be expressed as a line of sight integral of the matter over-density:

\begin{equation}
	g(\hat{\textbf{n}}) =\int_{0}^{\chi_{*}}d\chi\frac{H(\chi)}{c}W^{g}\delta(\chi\hat{\textbf{n}})
	\label{eq:galaxy_delta_relation}
\end{equation}
with $W^{g}$ given as
\begin{equation}
\begin{split}
	W^{g}(\chi) &= b\frac{H(\chi)}{c}\frac{dN}{dz(\chi)}+\frac{3\Omega_{m}}{2c^{2}}H_{0}^{2}(1+z)\chi\\ 
	&\times\int_{\chi}^{\chi_{*}}d\chi'\frac{H(\chi')}{c}\bigg(1-\frac{\chi}{\chi'}\bigg)(\alpha(\chi')-1)\frac{dN}{dz(\chi')}
\end{split}
	\label{eq:galaxy_kernel}
\end{equation}
where $\frac{dN}{dz}$ stands for the redshift distribution of galaxies,  $b$ is the galaxy linear bias that relates the luminous tracers of large scale structure with the underlying matter distribution and the second term accounts for the gravitational magnification of background objects by foreground sources (magnification bias; \citeauthor{Turner1980} \citeyear{Turner1980}). This effect depends on the slope, $\alpha(z)$, of the integral counts of sources above the flux threshold $S$, i.e. $N(>S)\propto S^{-\alpha}$. We estimate $\alpha$ as the slope of the straight line fit to $\log {N(>S)}$ distribution. Because the slope estimated for objects selected from the HELP catalogue used in this work is $\alpha=1$, the magnification bias term is null. However, in Section \ref{sec:discuss} we investigate the sensitivity of the results to different values of the slope.

\begin{figure}
    \centering
    \includegraphics[width=\linewidth]{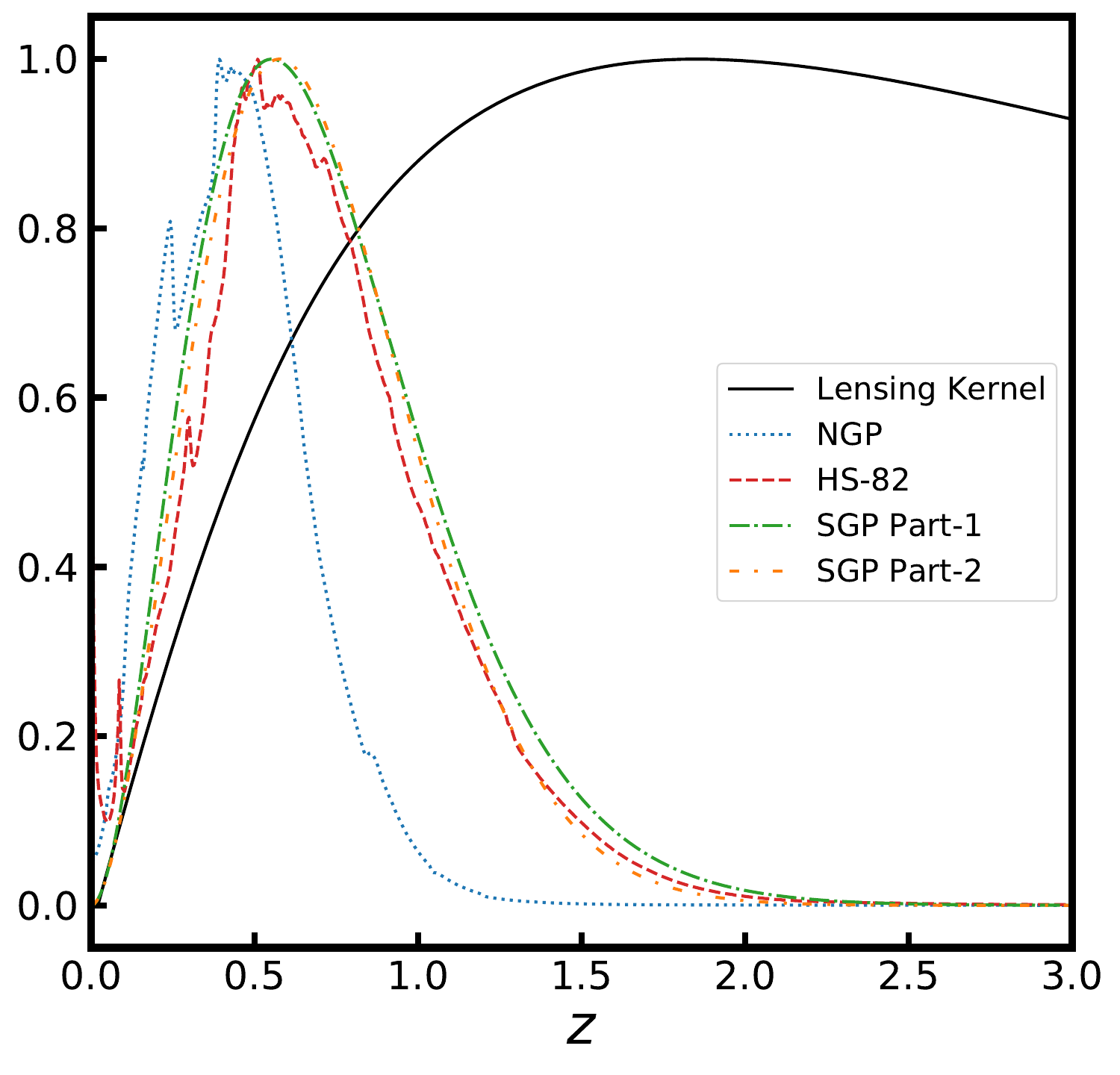}
    \caption{CMB lensing kernel $W^{\kappa}$ compared with redshift distributions for all galaxy patches. Both lensing kernel and redshift distributions are normalised to the unit maximum.}
    \label{fig:kernel_gal_dist}
\end{figure}

As we can see from Fig.~\ref{fig:kernel_gal_dist}, the lensing kernel is broad, slowly varying from $z\sim1$ and the redshift distributions of HELP catalogue galaxies peak at $z\sim 0.4$ for NGP, at $z\sim 0.5$ for HS-82 and SGP Part-1, and at $z\sim 0.6$ for SGP Part-2; the overlap between galaxy distribution peaks and lensing kernel ensures a high correlation between the galaxy over-density and CMB lensing convergence fields. The galaxy linear bias is, in general, a function of redshift $z$ and halo mass, $M$. Since the redshift distributions are broad, we assume a galaxy linear bias dependent on redshift via \mbox{$b(z) = b_{0}/D(z)$}, where $D(z)$ is the growth factor and $b_{0}$ is galaxy linear bias parameter.

The theoretical angular power spectrum is computed under the Limber approximation \citep{Limber1953} as

\begin{equation}
	C_{\ell}^{xy} = \int_{0}^{\chi_{*}}d\chi\frac{W^{x}(\chi)W^{y}(\chi)}{\chi^{2}}P(k=\frac{\ell+1/2}{\chi},z(\chi))
	\label{eq:power_spectra}
\end{equation} 
where $\{x,y\}=\{\kappa,g\}$, $\kappa\equiv$convergence and $g\equiv$galaxy over-density and $P(k=\frac{\ell+1/2}{\chi},z(\chi))$ is the matter power spectrum generated using CAMB\footnote{\url{https://camb.info/}}\citep{Lewis2000}. The mean redshift probed by the cross-correlation of CMB lensing convergence and the galaxy sample is given as:

\begin{equation}
	\langle z \rangle = \frac{\int_{0}^{\chi_{*}}d\chi z\frac{W^{\kappa}(\chi)W^{g}(\chi)}{\chi^{2}}P(k=\frac{\ell+1/2}{\chi},z(\chi))}{\int_{0}^{\chi_{*}}d\chi\frac{W^{\kappa}(\chi)W^{g}(\chi)}{\chi^{2}}P(k=\frac{\ell+1/2}{\chi},z(\chi))}
	\label{eq:mean_z}
\end{equation}
For our catalogue, the effective mean redshifts probed by the cross-correlation measurements are $\langle z \rangle = 0.60$, $0.77$, $0.76$ and $0.73$ for NGP, HS-82, SGP Part-1 and SGP Part-2, respectively.


\section{Data}\label{sec:data}

\subsection{CMB Lensing Data}\label{sec:planck_data}

The \textit{Planck} lensing data we use come from 2018 \textit{Planck} data release\footnote{\url{https://pla.esac.esa.int/\#cosmology}} described in \cite{Planck2020VIII}. It uses the SMICA DX12 CMB maps to reconstruct the lensing potential, covering $\sim 67$\% of the sky. For our baseline analysis, we use the lensing convergence map derived from a minimum-variance estimate of temperature and polarization data. We also use the CMB lensing potential map derived from the thermal Sunyaev-Zeldovich deprojected SMICA map for comparison of estimated parameters. Nevertheless, because of potentially smaller contamination by Cosmic Infrared Background (CIB) emission of the minimum-variance lensing map, compared to the lensing map derived from the Sunyaev-Zeldovich deprojected map (see discussion in section \ref{sec:cib}), we consider the former in our baseline analysis. For the rest of the paper, we dub the minimum-variance lensing convergence map as MV and Sunyaev-Zeldovich deprojected lensing convergence map as SZ-deproj. The lensing convergence which is proportional to the two dimensional Laplacian of the lensing potential (Eq. \ref{eq:2d_laplacian_relation}), can be expressed in spherical harmonic space using the relation \citep{Hu2000}

\begin{equation}
	\kappa_{\ell m} = -\frac{\ell(\ell+1)}{2}\phi_{\ell m}
	\label{eq:lensing_potential_convergence_relation}
\end{equation}

The spherical harmonic coefficients for lensing convergence are provided by \textit{Planck} data package in HEALPix\footnote{\url{https://healpix.jpl.nasa.gov/}} \citep{Gorski2005} format with $\textit{N}_{side}=4096$. As small angular scales are noise-dominated we transform these coefficients to a HEALPix map with lower resolution having $\textit{N}_{side}=512$, which we use for further analysis in this study. The data package also provides noise power spectra $N_{\ell}^{\kappa\kappa}$ for both MV and SZ-deproj maps, along with a binary map masking parts of the sky not used in the analysis.

\subsection{Galaxy Data}\label{sec:gal_data}

The \textit{Herschel} Extra-galactic Legacy Project (HELP\footnote{\url{https://herschel.sussex.ac.uk}}) is a catalogue combining data from 23 extra-galactic survey fields, observed between $0.36-4.5\,\mu m$. The catalogue covers $\sim 1300$ deg$^{2}$ and contains objects with cross-matching between 51 public surveys \citep{Raphael2019,Raphael2021}. It is worth pointing out that in our analysis we use only data from these public surveys covering \textit{Herschel} fields, but not far-infrared data from the \textit{Herschel} satellite itself as was done previously in similar studies \citep{Bianchini2015,Bianchini2016}. Most of the extra-galactic fields available with HELP have a very small physical surface area, making them unsuitable for our cross-correlation study. Also, because different surveys have different depths, some fields are inhomogeneous and cannot be used for our analysis. Combining these constraints we are left with three useful patches for our analysis: North Galactic Pole, South Galactic Pole, and \textit{Herschel}-Stripe 82 (hereafter, NGP, SGP, and HS-82, respectively) covering a total of $\sim 660$ deg$^{2}$. The details of the properties associated with each patch on the sky are provided in Table \ref{tab:HELP_data} while in Table \ref{apndx_tab:bandwise_coverage_help_fields} is shown the percentage of objects for a given field observed by specific surveys and their photometric filters.

\begin{table*}
	\centering
	\captionsetup{justification=centering}
	\caption{Physical properties of HELP patches. $[l,b]$ are galactic longitude and latitude respectively, $N_{obj}$ is the number of objects in each patch, $\overline{n}$ is the mean number of objects, $f_{sky}$ is the fraction of sky covered by patches, and $\langle z \rangle$ is the mean redshift probed by cross-correlation computed from Eq. \ref{eq:mean_z}.}
	\label{tab:HELP_data}
	\begin{tabular}{lcccccccc} 
		\hline\hline
		Patch & $f_{sky}$ & area [deg$^{2}$] & [$l$,$b$] & $N_{obj}$ & $\overline{n}$ [gal pix$^{-1}$] & $\overline{n}$ [gal str$^{-1}$] & median $z$ & $\langle z \rangle$\\
		\hline
		NGP & 0.0043 & 179.14 & [51$^{\degr}$,84$^{\degr}$] & 1311549 & 96.908 & 2.426$\times$10$^{7}$ & 0.45 & 0.60\\
		HS-82 & 0.0062 & 255.16 & [130$^{\degr}$,-61$^{\degr}$] & 6824474 & 344.862 & 8.633$\times$10$^{7}$ & 0.60 & 0.77\\		
		SGP Part-1 & 0.0020 & 85.83 & [12$^{\degr}$,-68$^{\degr}$] & 3151922 & 481.577 & 1.206$\times$10$^{8}$ & 0.71 & 0.76\\
		SGP Part-2 & 0.0035 & 145.32 & [-82$^{\degr}$,-86$^{\degr}$] & 6659404 & 600.975 & 1.504$\times$10$^{8}$ & 0.71 & 0.73\\
		\hline
	\end{tabular}
\end{table*}
The galaxy density of the SGP field has significant variations due to the fact that the Kilo-Degree Survey (KiDS; \citealt{Kids2013}) only covers approximately one-half of the SGP field, whereas the Dark Energy Survey (DES; \citealt{DESY12018}) covers the other half. These two halves have a large difference in their mean number of objects per pixel. To avoid any effects coming from this inhomogeneity, we divide the SGP field into two parts, which we refer to as SGP Part-1 and SGP Part-2. 

To remove star-like sources and increase the purity of the galaxy sample, we apply two extra selection criteria, \texttt{flag-gaia}$\leq 2$ and \texttt{stellarity}$<0.9$, which increase the likelihood that an object is a galaxy.

HELP uses the Easy and Accurate Z from Yale code (EAZY; \citeauthor{Brammer2008} \citeyear{Brammer2008}) to estimate photo-$z$ for objects. It provides posteriors from the photo-$z$ pipeline for every object in the catalogue. The estimated photometric redshifts are the median values of these posteriors. We use these posteriors to build on the redshift galaxy distributions by stacking these posteriors together. The redshift distributions are shown in Fig. \ref{fig:kernel_gal_dist}. Using these posteriors translates the errors on redshift to redshift distributions \citep{Budavari2003}. We select all objects for which the relative error on redshift, $\frac{\sigma_{z}}{1+z}< 0.15$ for NGP and HS-82 and $\frac{\sigma_{z}}{1+z}< 0.25$ for SGP Part-1 and SGP Part-2. We use a higher limit on $\frac{\sigma_{z}}{1+z}$ for SGP fields, because 
the median value of $\frac{\sigma_{z}}{1+z}$ for these fields is higher, i.e.~0.22, than for NGP and HS-82 fields which have the median equal to 0.15 and 0.16, respectively.

These cuts result in a final catalogue of $\sim 18$ million objects. Table \ref{tab:HELP_data} summarises the total and mean number of objects for the NGP, HS-82, and the two SGP fields. The galaxy number density of the samples used in this analysis is approximately two orders of magnitudes higher than that of the far-infrared Herschel data set used in \citeauthor{Bianchini2015} (\citeyear{Bianchini2015, Bianchini2016}).

\begin{figure*}
    \begin{subfigure}{.25\linewidth}
        \centering
        \includegraphics[width=4cm]{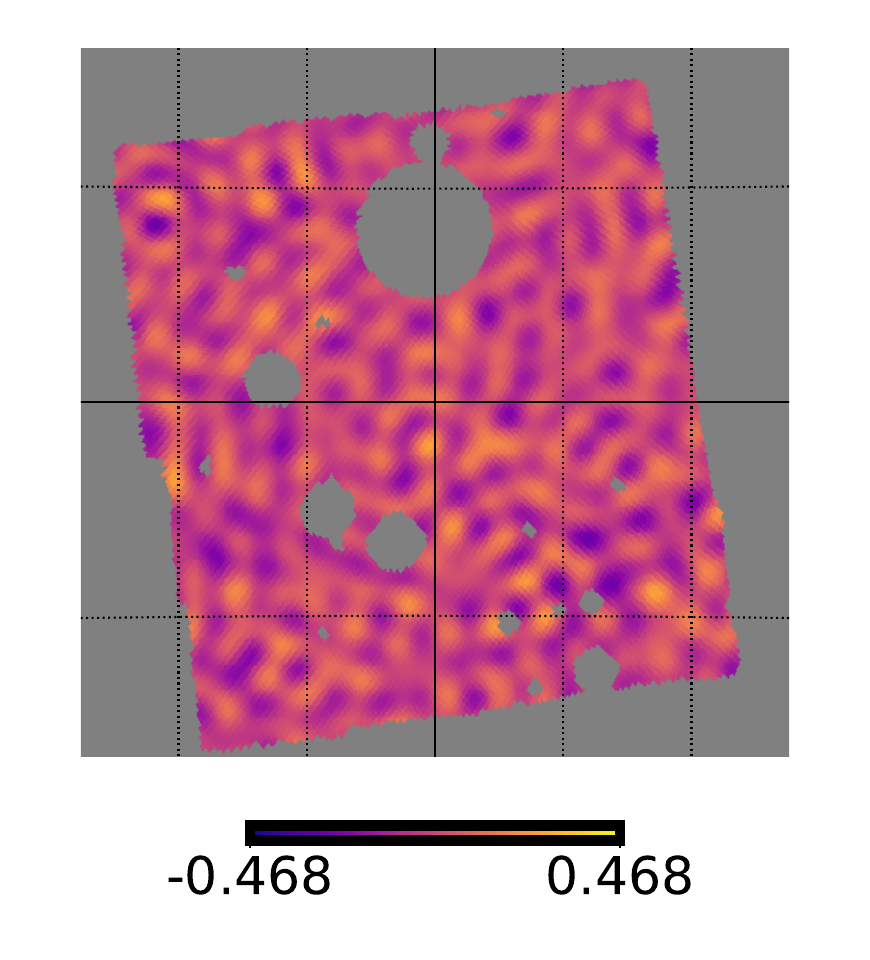}
        \captionsetup{labelformat=empty}
        \caption{NGP}
    \end{subfigure}%
    \begin{subfigure}{.25\linewidth}
        \centering
        \includegraphics[width=4cm]{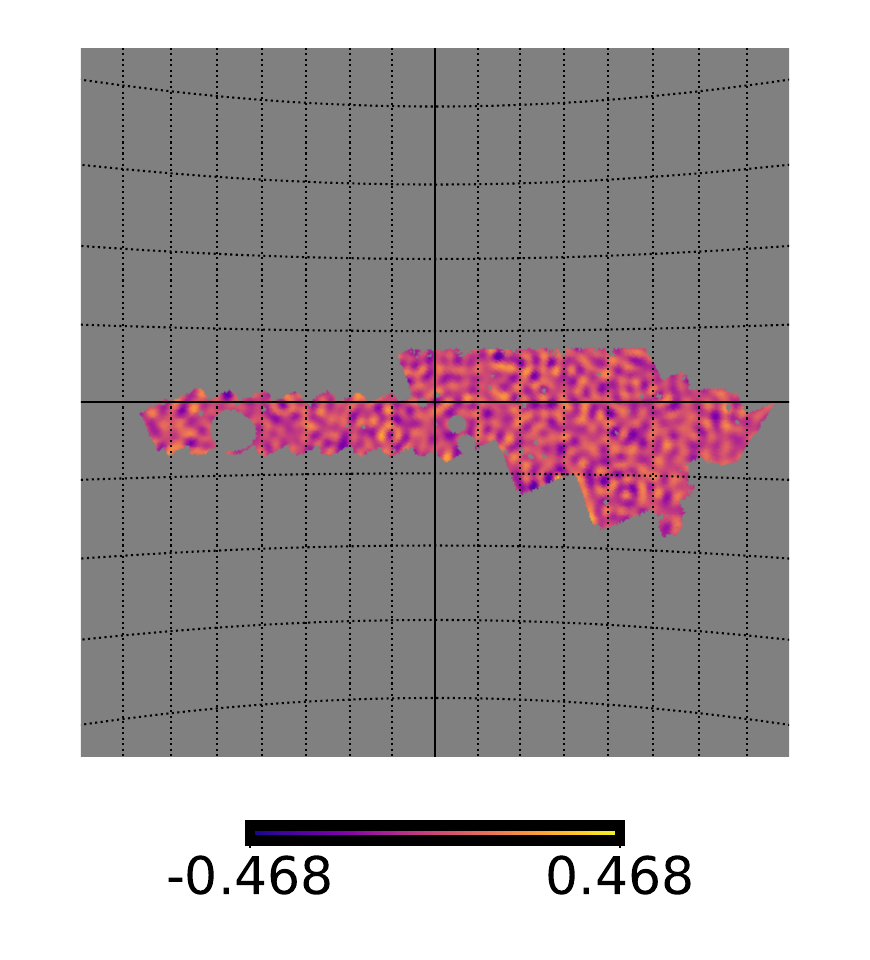}
        \captionsetup{labelformat=empty}
        \caption{HS-82}
    \end{subfigure}%
    \begin{subfigure}{.25\linewidth}
        \centering
       \includegraphics[width=4cm]{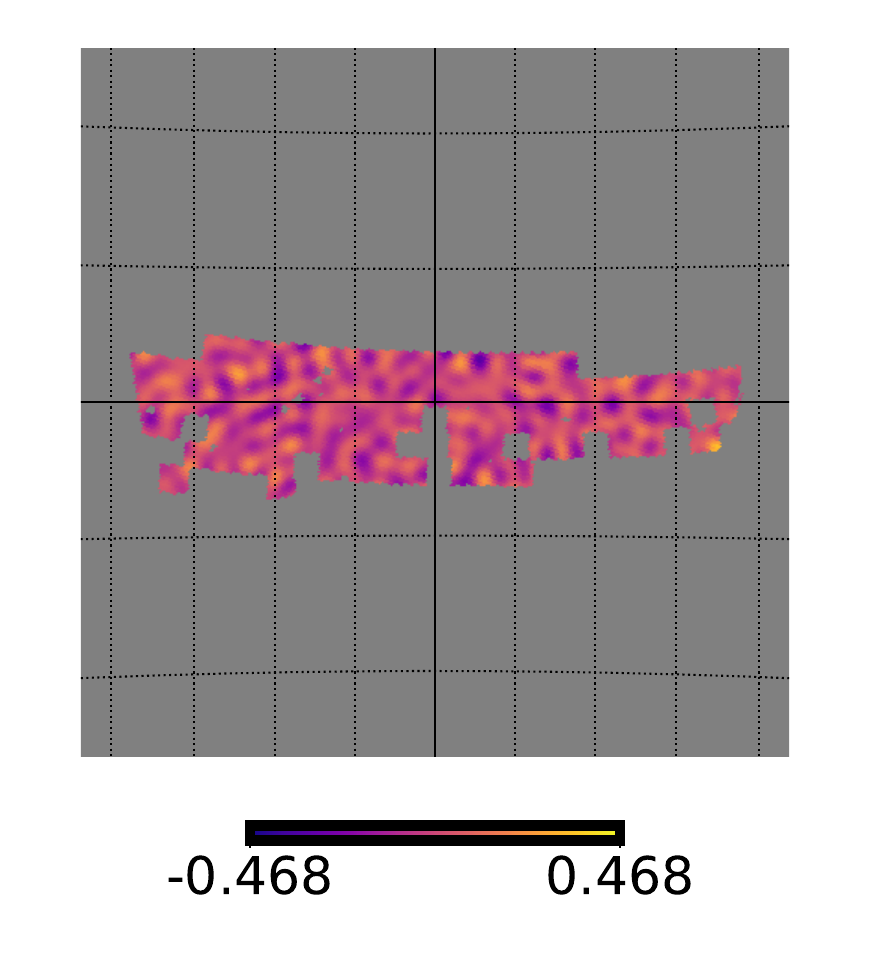}
       \captionsetup{labelformat=empty}
        \caption{SGP Part-1}
    \end{subfigure}%
    \begin{subfigure}{.25\linewidth}
        \centering
       \includegraphics[width=4cm]{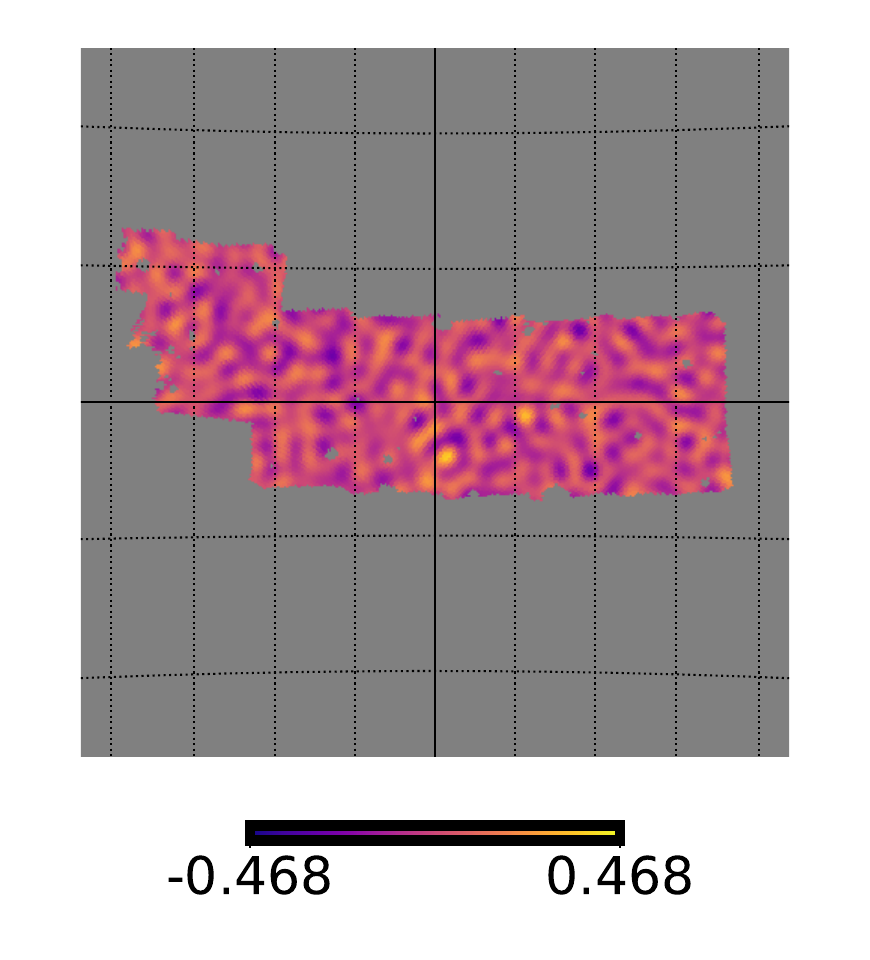}
       \captionsetup{labelformat=empty}
        \caption{SGP Part-2}
    \end{subfigure}\\[1ex]
    \begin{subfigure}{.25\linewidth}
        \centering
        \includegraphics[width=4cm]{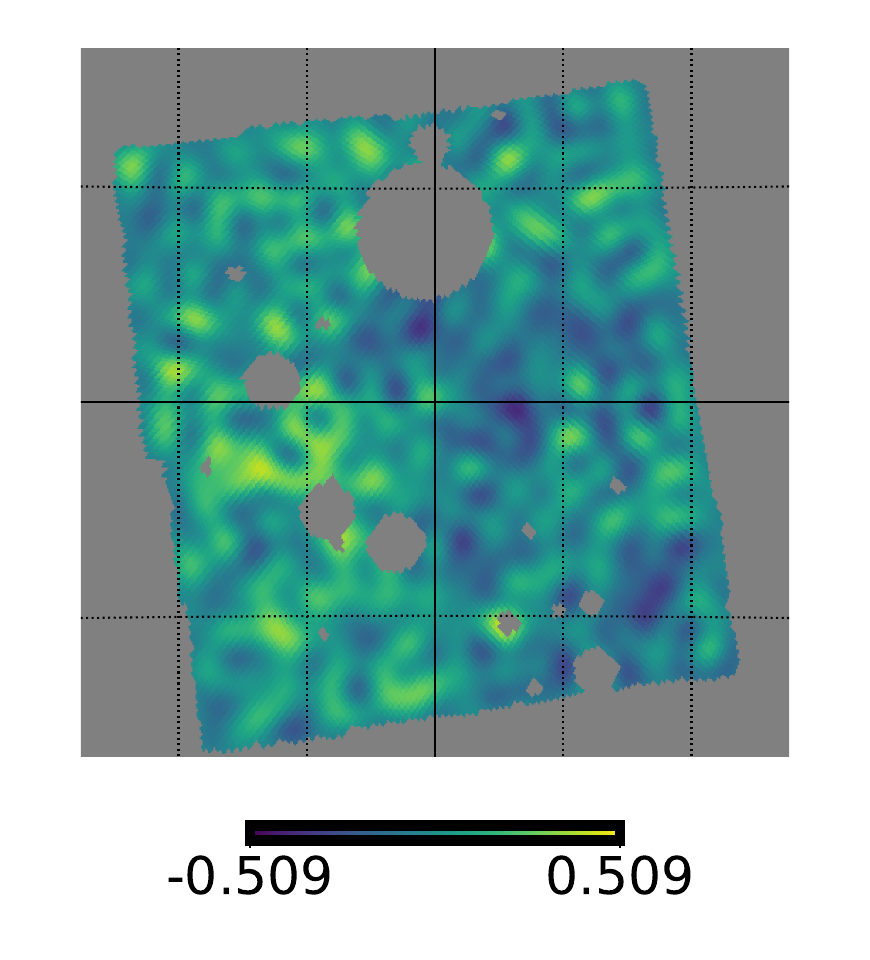}
        \captionsetup{labelformat=empty}
        \caption{NGP} 
    \end{subfigure}%
    \begin{subfigure}{0.25\linewidth}
        \centering
        \includegraphics[width=4cm]{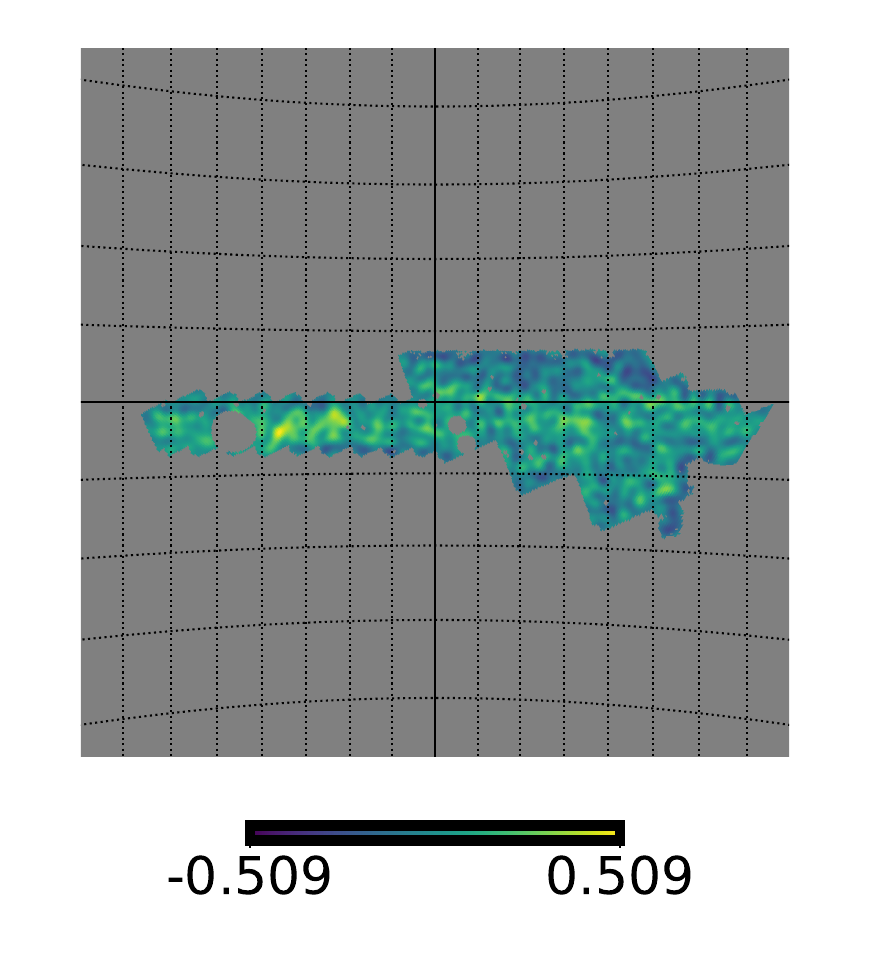}
        \captionsetup{labelformat=empty}
        \caption{HS-82} 
    \end{subfigure}%
    \begin{subfigure}{.25\linewidth}
        \centering
        \includegraphics[width=4cm]{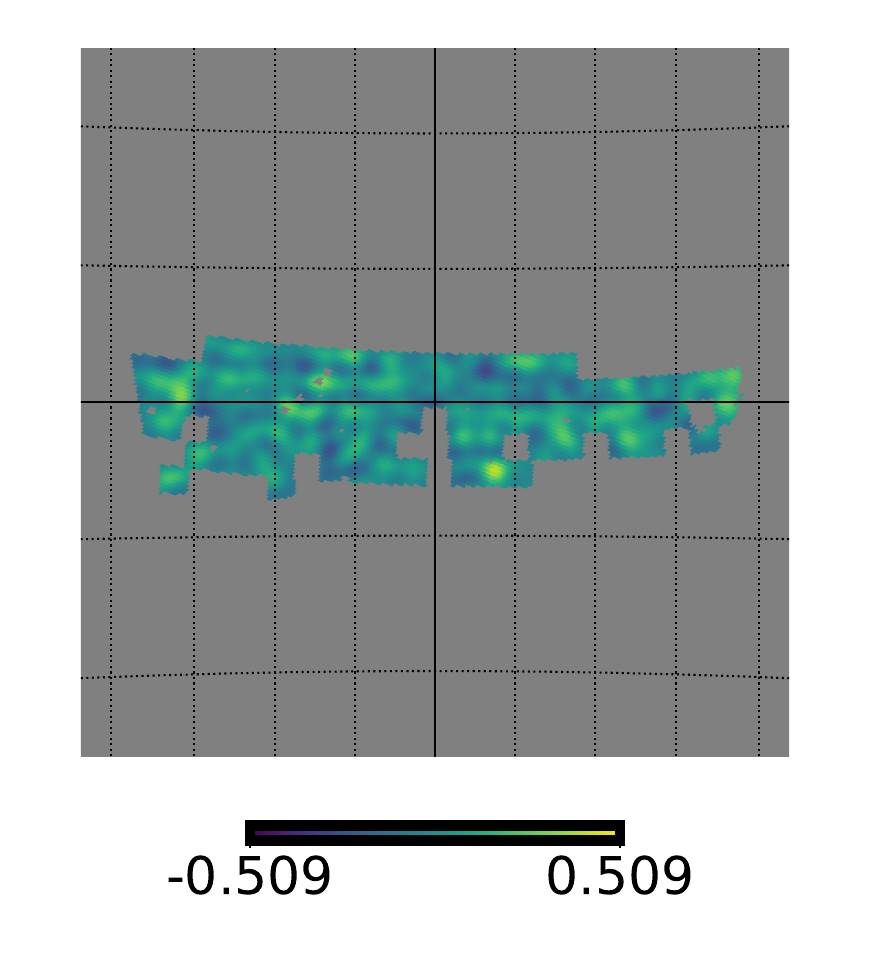}
        \captionsetup{labelformat=empty}
        \caption{SGP Part-1}
    \end{subfigure}%
    \begin{subfigure}{.25\linewidth}
        \centering
        \includegraphics[width=4cm]{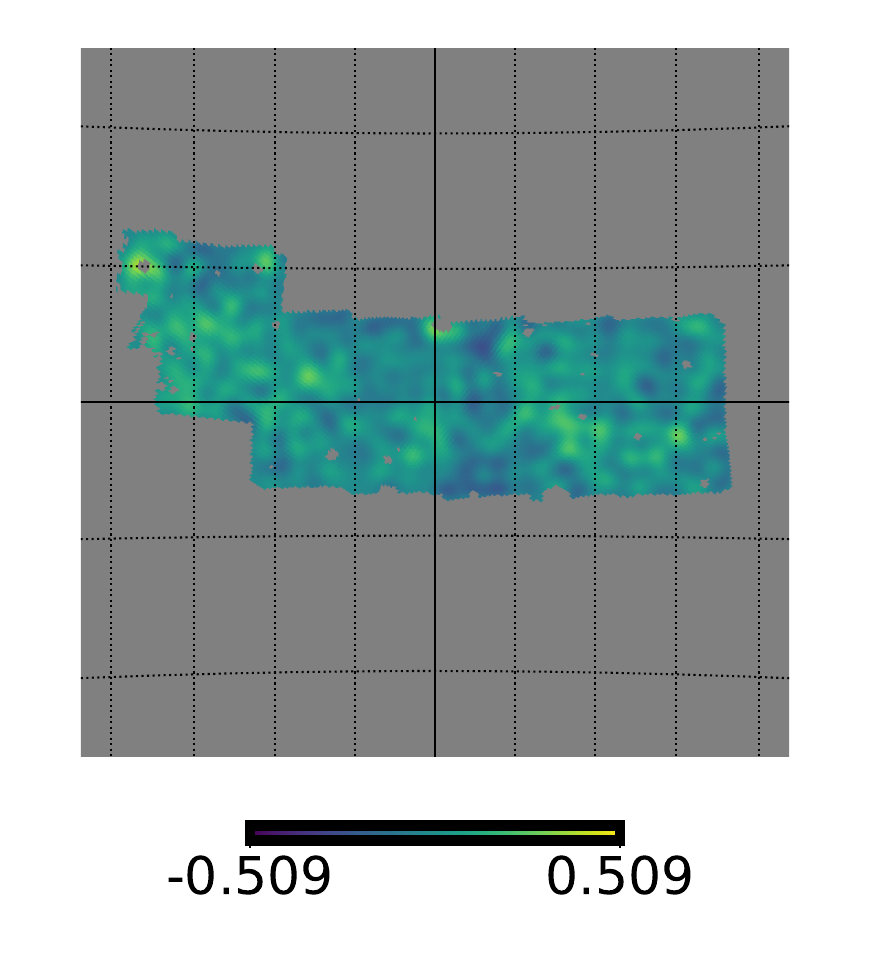}
        \captionsetup{labelformat=empty}
        \caption{SGP Part-2}
    \end{subfigure}
    \caption{Convergence maps (upper row) and galaxy over-density Maps (lower row) of NGP, HS-82, SGP Part-1 and SGP Part-2 fields. Multipoles $\ell\geq 400$ have been filtered out from all maps. The grid spacing is $3^{\degr}$ in longitude and $5^{\degr}$ in latitude.}
    \label{fig:filtered_maps}
\end{figure*}

We build galaxy over-density maps with resolution $\textit{N}_{side}=512$ using the relation

\begin{equation}
	g(\hat{\textbf{n}}) = \frac{n(\hat{\textbf{n}})-\overline{n}}{\overline{n}}
	\label{eq:gal_overdensity}
\end{equation}
where $n(\hat{\textbf{n}})$ is the number of objects in a given pixel and $\overline{n}$ is the mean number of objects per pixel. Fig. \ref{fig:filtered_maps} shows the \textit{Planck} lensing convergence and galaxy over-density maps for all patches, from which we have filtered out $\ell\geq 400$. It shows the part of the sky common to both galaxy fields and \textit{Planck} convergence field. But in our analysis, we use the full lensing convergence map available (covering $\sim 67\,\%$), uplifting the condition of using only the common area.

\section{Methodology}\label{sec:methodology}

\subsection{Power-spectra}\label{sec:power_spectra}

We use the MASTER algorithm \citep{Hivon2002} to convert from pseudo-power spectra computed from smaller fractions of the sky to the full-sky power spectra. MASTER is based on direct spherical harmonic transform to obtain an unbiased estimate of the power spectrum. The harmonic mode coupling induced by incomplete sky coverage is described by the mode coupling kernel $M_{\ell\ell}$. The ensemble average of the pseudo-power spectrum can be related to the full-sky power spectrum by

\begin{equation}
	\tilde{C}_{\ell}^{xy} = \sum_{\ell '}M_{\ell\ell '}C_{\ell '}^{xy}
	\label{eq:pseudo_full_relation}
\end{equation}
where $C_{\ell}^{xy}$ is the full-sky power spectrum and $\tilde{C}_{\ell}^{xy}$ is the pseudo-power spectrum measured from data from the relation
\begin{equation}
    \tilde{C}_{\ell}^{xy} = \frac{1}{2\ell+1}\sum\limits_{m=-\ell}^{\ell}\tilde{a}_{\ell m}^{x}\tilde{a}_{\ell m}^{*y}
    \label{eq:pseudo_cl_alm_relation}
\end{equation}
where $\tilde{a}_{\ell m}$ are the spherical harmonic coefficients from partial sky coverage and $\{x,y\} = \{\kappa,g\}$.

However, Eq.~\ref{eq:pseudo_full_relation} cannot be inverted directly because the coupling kernel is singular for smaller fractions of sky. To avoid these singularities  we bin the power spectrum in $\ell$ with bin-width of $\Delta\ell = 100$ and the estimator of power spectra is then given by

\begin{equation}
	\langle \hat{C}_{L}^{xy} \rangle = \sum_{L'}K_{LL'}^{-1}(\langle\tilde{C}_{L'}^{xy}\rangle-\langle\tilde{N}_{L'}^{xy}\rangle)
	\label{eq:pseudo_full_relation_bin}
\end{equation}
where
\begin{equation}
	\tilde{C}_{L'}^{xy} = \sum_{\ell'}P_{L'\ell'}\tilde{C}_{\ell'}^{xy}
	\label{eq:binned_spectra}
\end{equation}
and
\begin{equation}
	K_{LL'} = \sum_{\ell\ell'}P_{L\ell}M_{\ell\ell '}B_{\ell'}^{2}Q_{\ell'L'}
	\label{eq:binned_kernel}
\end{equation}
where $L$ stands for the multipole bin and $B_{\ell}$ is the pixel window function that accounts for the finite size of the pixels. $P_{L\ell}$ is the binning operator expressed as:
\begin{equation}
   P_{L\ell} = 
   \begin{cases}
   \frac{1}{2\pi}\frac{\ell(\ell+1)}{\ell_{\text{low}}^{(L+1)}-\ell_{\text{low}}^{(L)}},& \text{if}\quad 2\leq\ell_{\text{low}}^{(L)}\leq\ell\leq\ell_{\text{low}}^{(L+1)}\\
   0,& \text{otherwise}
   \end{cases}
   \label{eq:binning_operator}
\end{equation}
and $Q_{\ell L}$ is the reciprocal binning operator:
\begin{equation}
    Q_{\ell L} = 
    \begin{cases}
        \frac{2\pi}{\ell(\ell+1)},& \text{if}\quad 2\leq\ell_{\text{low}}^{(L)}\leq\ell\leq\ell_{\text{low}}^{(L+1)}\\
        0,& \text{otherwise}
    \end{cases}
    \label{eq:reciprocal_binning_operator}
\end{equation}

In Eq.~\ref{eq:pseudo_full_relation_bin}, $\tilde{N}_{L}^{xy}$ is the pseudo noise power spectrum. In cross-correlation studies, one of the main assumption is that the noise associated with galaxy over-density and lensing convergence fields are uncorrelated. Thus, there will be no noise associated with cross-power spectrum, i.e., $N_{\ell}^{\kappa g} = 0$. But for auto-power spectra $C_{\ell}^{\kappa\kappa}$ and $C_{\ell}^{gg}$, we account for noise. We estimate the noise pseudo spectra from Monte Carlo simulations, with which Eq. \ref{eq:pseudo_full_relation_bin} becomes
\begin{equation}
	\hat{C}_{L}^{xy} = \sum_{L'}K_{LL'}^{-1}(\tilde{C}_{L'}^{xy}-\langle\tilde{N}_{L'}^{xy}\rangle_{MC})
	\label{eq:pseudo_full_relation_bin_L}
\end{equation}
We estimate the full-sky power spectrum for the multipole range $0\leq \ell \leq 1200$ using Eq.~\ref{eq:pseudo_full_relation_bin_L}, while we use $7$ linear bins in the multipole range $100\leq \ell \leq 800$  in our study.

\subsection{Errors}\label{sec:errors}

The errors on the power spectrum can be computed from the square root of diagonal of the analytical covariance matrix:

\begin{equation}
	\begin{split}
	Cov_{LL'}^{AB,CD} = &\frac{1}{(2\ell_{L'}+1)\Delta\ell f_{sky}^{AB}f_{sky}^{CD}}\bigg[f_{sky}^{AC,BD}\sqrt{C_{L}^{AC}C_{L'}^{AC}C_{L}^{BD}C_{L'}^{BD}}\\
&+f_{sky}^{AD,BC}\sqrt{C_{L}^{AD}C_{L'}^{AD}C_{L}^{BC}C_{L'}^{BC}}\bigg]\delta_{LL'}
	\end{split}
	\label{eq:error_covariance}
\end{equation}
where $\{A,B,C,D\}=\{\kappa,g\}$, $\Delta\ell$ is the multipole binwidth, $f_{sky}^{AB}$ is the fraction of sky common to fields $A$ and $B$, $f^{AC;BD}$ is the fraction of sky covered by fields $AB$ and $CD$, and $\delta_{LL'}$ is the Kronecker delta. The expression of covariance matrix in Eq.~\ref{eq:error_covariance} is not constrained by the limitation of using only the fraction of sky common to both CMB lensing convergence and galaxy surveys. The detailed derivation of the covariance matrix can be found in appendix \ref{apndx:covariance_matrix}. We use the expression Eq.~\ref{eq:error_covariance} for covariance in estimation of parameters from likelihood analysis discussed in section \ref{sec:likeli}.

\begin{figure}
    \centering
    \includegraphics[width=\linewidth]{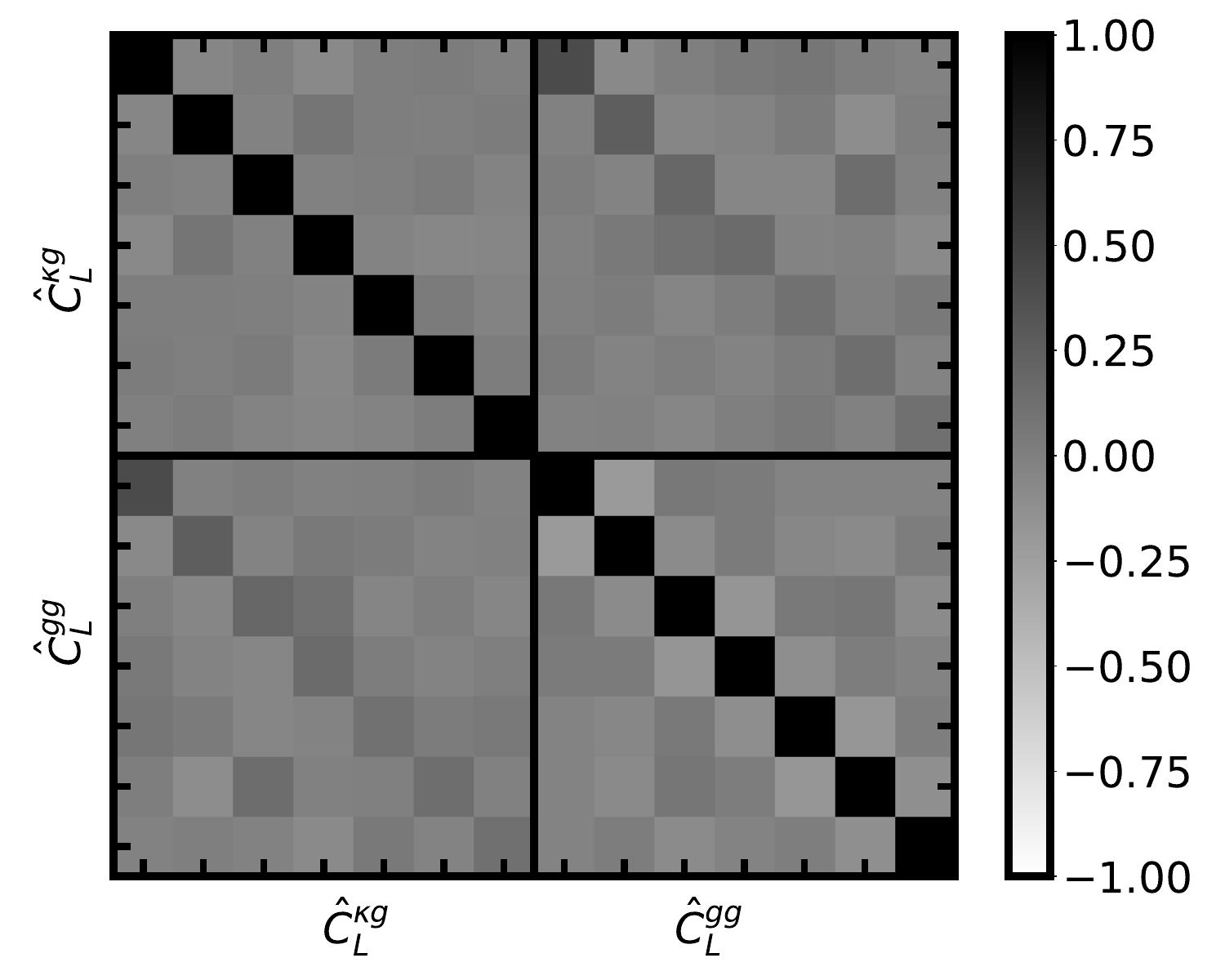}
    \caption{Correlation matrices constructed from the covariance matrix mentioned in Eq. \ref{eq:cov_simul} for HS-82 field. The correlation matrix is dominantly diagonal.}
    \label{fig:correlation_matrices}
\end{figure}
Fig. \ref{fig:correlation_matrices} shows the correlation matrices obtained from simulations of HS-82 field. All correlation matrices are dominated by diagonal elements and the off-diagonal terms are smaller in comparison. Thus the assumed diagonal approximation holds strongly for our purpose and also serves as a good way to save some computation time.

\subsection{Simulations}
\label{sec:simulations}
To validate the numerical algorithms and to check that power-spectra and parameters estimated do not contain any systematic error, we simulate maps of both CMB convergence and galaxy over-density fields with statistical properties consistent with observations. We introduce a known degree of correlation in theoretical power-spectra (Eq.~\ref{eq:power_spectra}), using the relation \citep{Kamionkowski1997}:

\begin{equation}
\begin{aligned}
\kappa_{\ell m} &= \xi_{1}(C_{\ell}^{\kappa\kappa})^{1/2}; \\
g_{\ell m} &= \xi_{1}\frac{C_{\ell}^{\kappa g}}{(C_{\ell}^{\kappa\kappa})^{1/2}}+\xi_{2}\bigg[C_{\ell}^{gg}-\frac{(C_{\ell}^{\kappa g})^{2}}{C_{\ell}^{\kappa\kappa}}\bigg]^{1/2}
\end{aligned}
\label{eq:simuleqn}
\end{equation}

For each $\ell$ and $m > 0$, $\xi_{1}$ and $\xi_{2}$ are two complex random numbers drawn from a Gaussian distribution with unit variance and for $m=0$, the random numbers are real and normally distributed.\\

We simulate these maps to also include noise. For convergence noise maps, we used the minimum variance noise power spectrum $N_{\ell}^{\kappa\kappa}$, provided in the \textit{Planck} 2018 data package. To account for noise associated with galaxy over-density maps, we simulate galaxy number count maps where the value in each pixel is drawn from a Poisson distribution with mean

\begin{equation}
	\lambda(\hat{\textbf{n}}) = \overline{n}(1+g(\hat{\textbf{n}}))
	\label{eq:Poisson_noise}
\end{equation} 
where $\overline{n}$ is the mean number of sources per pixel and $g(\hat{\textbf{n}})$ is the corresponding simulated galaxy over-density map with only signal. When simulating Gaussian fields, it often leads to some pixels with values $g<-1$ from which Poisson samples cannot be drawn. In cases of simulations with some pixels being $g<-1$, we reject those simulated maps. The galaxy number count map, thus obtained, is then converted to the over-density map using Eq.~\ref{eq:gal_overdensity}. The statistical properties of noise simulated in this way are the same as for data.

From maps simulated with this procedure, we recover the full sky power-spectra as mentioned in section \ref{sec:power_spectra}. The mean power spectrum is computed as

\begin{equation}
	 \overline{C}_{L}^{xy}\equiv\langle\hat{C}_{L}^{xy}\rangle = \frac{1}{N}\sum_{i=1}^{N}\hat{C}_{L}^{xy,i}
	\label{eq:average_spectra}
\end{equation}
where $\hat{C}_{L}^{xy,i}$ represents the power spectrum estimate for $i^{th}$ simulation and $N$ is the number of simulations. The associated errors are computed from the covariance matrix as

\begin{equation}
	\Delta \overline{C}_{L}^{xy} = \bigg(\frac{Cov_{LL}^{xy}}{N}\bigg)^{1/2}
	\label{eq:err_simul}
\end{equation}
where the covariance matrix is evaluated from simulations:

\begin{equation}
	Cov_{LL'}^{xy} = \frac{1}{N-1}\sum_{i=1}^{N}(\hat{C}_{L}^{xy,i}-\overline{C}_{L}^{xy})(\hat{C}_{L'}^{xy,i}-\overline{C}_{L'}^{xy})
	\label{eq:cov_simul}
\end{equation}


\section{Likelihood Analysis and Parameter Estimation}\label{sec:likeli}

We estimate two parameters in our study: galaxy linear bias parameter $b_{0}$ and amplitude of the cross-power spectrum, $A$. The cross-power spectrum depends on the product of galaxy linear bias parameter $b_{0}$ and amplitude $A$. To break this degeneracy, we use a likelihood function containing galaxy auto-power spectrum, which scales as $b_{0}^{2}$ as well as cross-power spectrum. This joint likelihood function is given as:

\begin{equation}
\begin{split}
	\mathcal{L}&(\hat{C}_{L}|b_{0},A) = \frac{1}{\sqrt{(2\pi)^{N_{L}}det(Cov_{LL'})}} \times\\
	& \times \text{exp}\bigg\lbrace -\frac{1}{2}[\hat{C}_{L}-C_{L}(b_{0},A)](Cov_{LL'})^{-1}[\hat{C}_{L'}-C_{L'}(b_{0},A)]\bigg\rbrace
\end{split}
\label{eq:joint_likeli}
\end{equation}
where $N_{L}$ is the number of multipole bins. $\hat{C}_{L}$ is 14-element data vector constructed from estimated galaxy auto-power spectrum $\hat{C}_{L}^{gg}$ and cross-power spectrum $\hat{C}_{L}^{\kappa g}$ as

\begin{equation}
	\hat{C}_{L} = (\hat{C}_{L}^{\kappa g},\hat{C}_{L}^{gg})
	\label{eq:joint_vector}
\end{equation}
$C_{L}(b_{0},A)$ is the joint power spectrum template, used in extracting parameters $b_{0}$ and $A$, defined as

\begin{equation}
    C_{L}(b_{0},A) = (AC_{L}^{\kappa g}(b_{0}),C_{L}^{gg}(b_{0}))
\end{equation}
The covariance matrix $Cov_{LL'}$ in Eq. \ref{eq:joint_likeli} is given as

\begin{equation}
	Cov_{LL'} = 
	\begin{bmatrix}
		Cov_{LL'}^{\kappa g,\kappa g} & Cov_{LL'}^{\kappa g,gg} \\
		Cov_{LL'}^{\kappa g,gg} & Cov_{LL'}^{gg,gg}	
	\end{bmatrix}
	\label{eq:joint_cov_full}
\end{equation}
where the covariance $Cov_{LL'}^{\kappa g,gg}$ which accounts for the correlation between the CMB convergence and galaxy density fields, is given by Eq.~\ref{eq:error_covariance} with $(A,B,C,D)\equiv (\kappa,g,g,g)$. Similarly, expressions for $Cov_{LL'}^{\kappa g}$ and $Cov_{LL'}^{gg}$ is evaluated using $(A,B,C,D)\equiv (\kappa,g,\kappa,g)$ and $(A,B,C,D)\equiv (g,g,g,g)$, respectively.

We have two free parameters that we estimate from likelihood analysis namely, galaxy linear bias $b_{0}$ and cross-correlation amplitude $A$. We use flat priors for these parameters with $\mathbf{b_{0} \in [0,10]}$ and $\mathbf{A \in [-5,5]}$. The remaining cosmological parameters are kept constant with values from our fiducial background cosmology described in \cite{Planck2020VI} (i.e.~the flat $\Lambda$CDM cosmology with the best-fit \textit{Planck} + \textit{WP} + highL + lensing parameters). We use the publicly available package EMCEE \citep{EMCEE2013} to sample the parameter space. We define the best-fit value of each parameter to be the median of the posterior distribution after marginalizing over the other free parameter. We additionally report the $1\,\sigma$ confidence intervals of the estimated best-fit values.


\section{Pipeline Validation}
\label{sec:validation}
\begin{figure*}
    \begin{subfigure}[b]{0.3\linewidth}
        \centering
        \includegraphics[width=\linewidth]{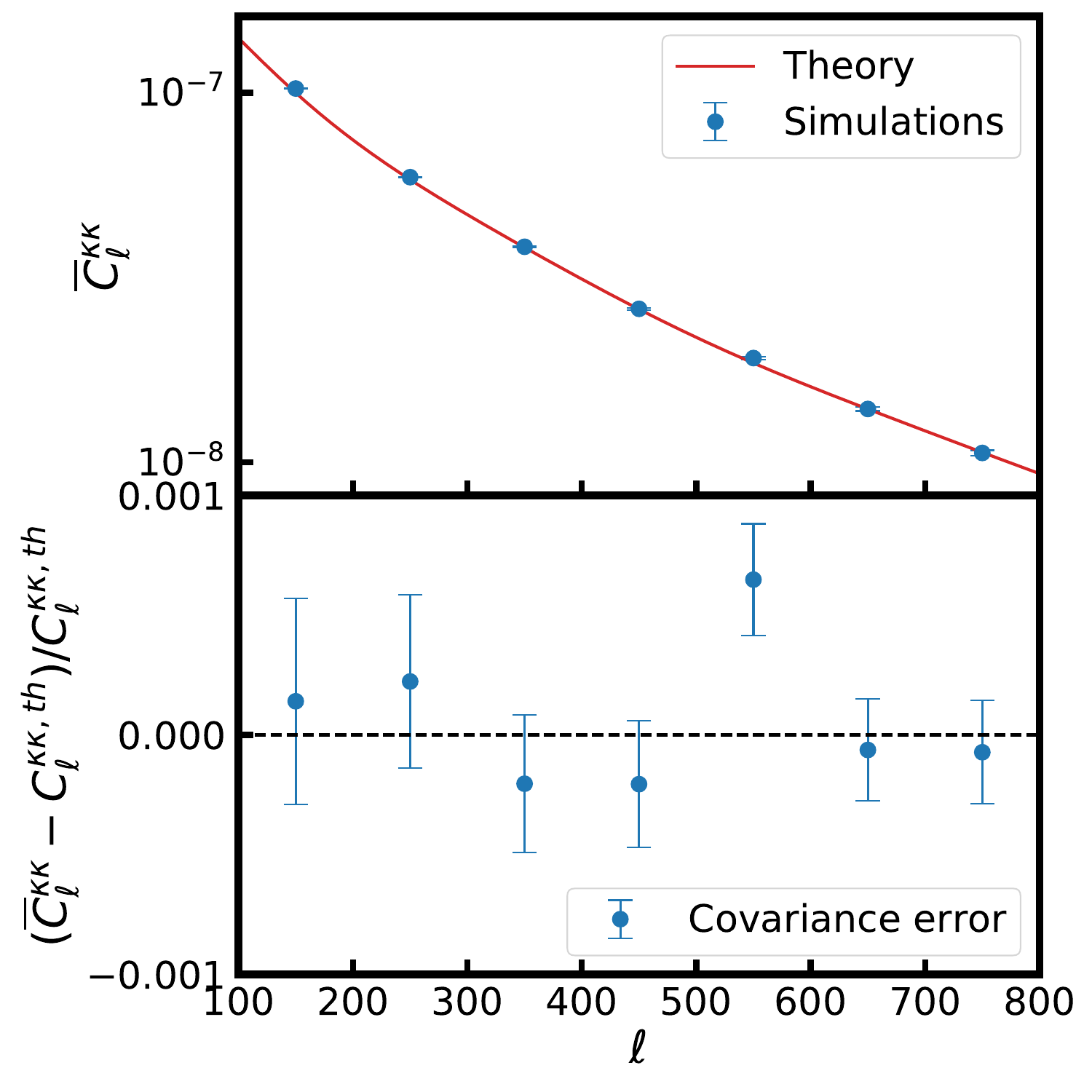}
    \end{subfigure}%
    \begin{subfigure}[b]{0.3\linewidth}
        \centering
        \includegraphics[width=\linewidth]{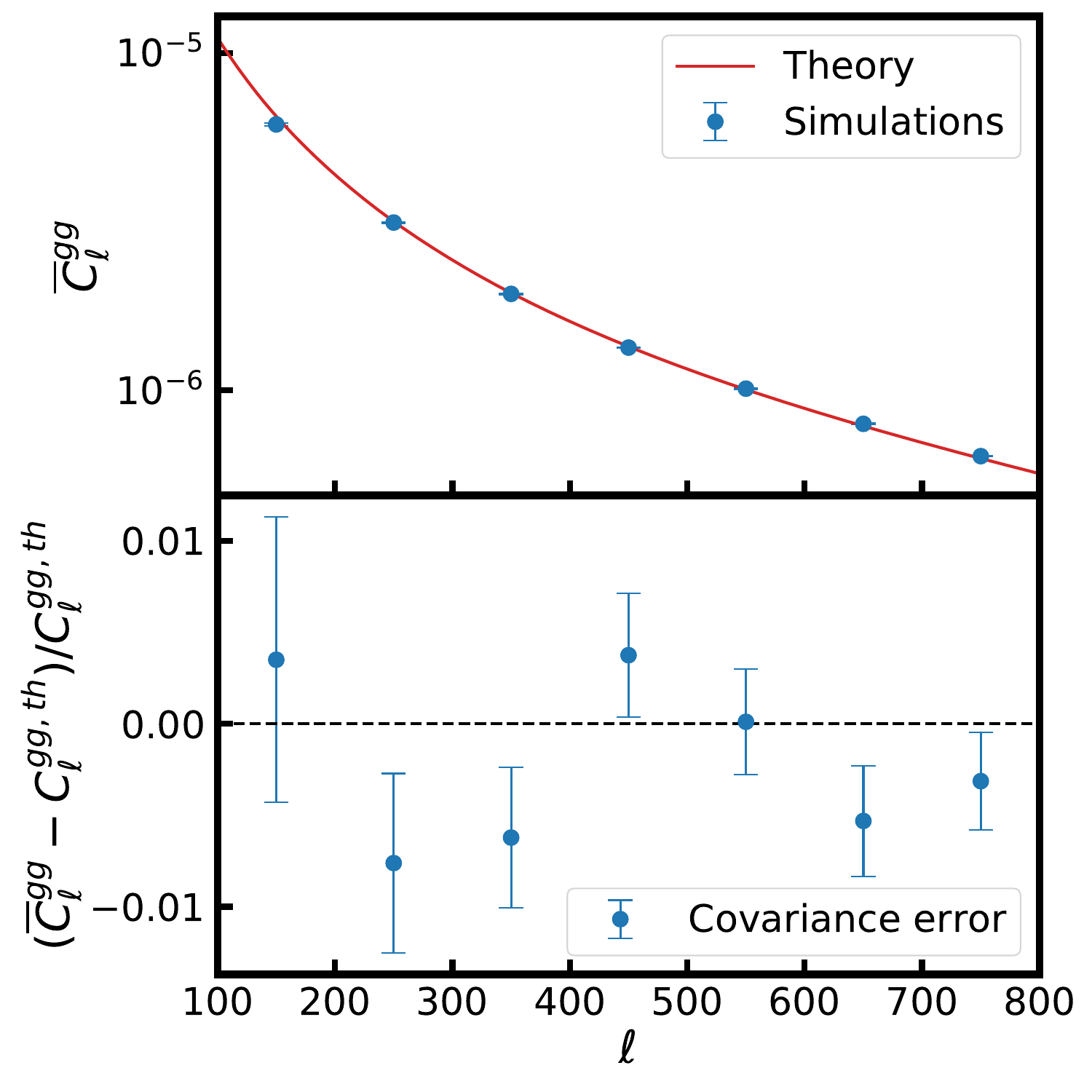}
    \end{subfigure}%
    \begin{subfigure}[b]{0.3\linewidth}
        \centering
        \includegraphics[width=\linewidth]{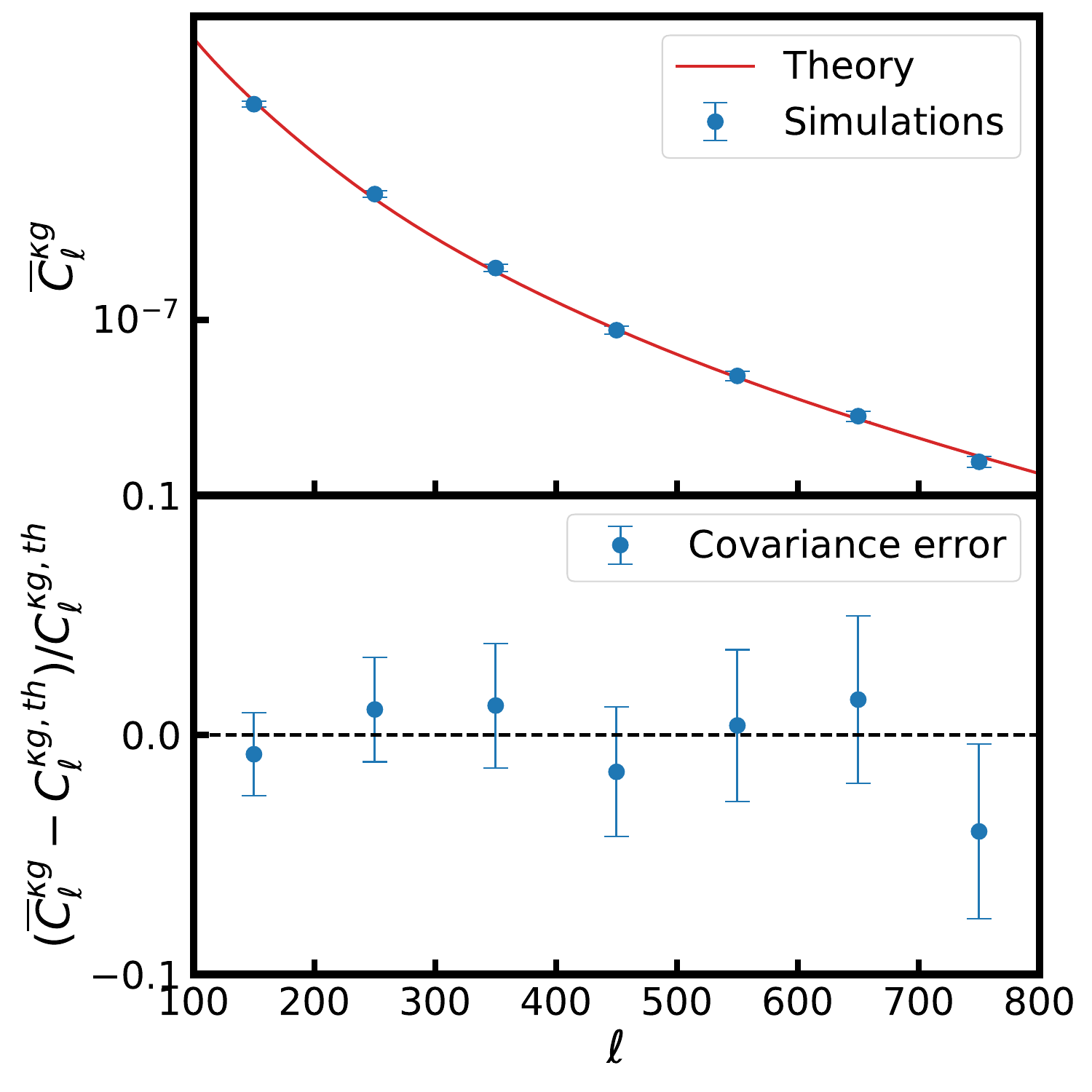}
    \end{subfigure}
    \caption{\textit{Top} : Average power spectra reconstructed for 500 simulations of HS-82 field. The red line represents the theoretical power spectrum used for simulations. \textit{Bottom}: Error estimated for the reconstructed simulated average power spectra relative to the theoretical power spectrum. \textit{Left to Right}: CMB lensing convergence power spectrum, galaxy auto-power spectrum, and cross-power spectrum}
    \label{fig:power_spectra_simulated}
\end{figure*}

We simulate 500 maps for HS-82 field with the procedure outlined in section \ref{sec:simulations} to validate our pipeline for estimating power spectra and parameters. The upper panel of Fig.~\ref{fig:power_spectra_simulated} shows the reconstructed power spectra averaged over 500 simulations using our algorithm. The lower panel shows the relative difference between the theoretical power spectrum used for simulations and the recovered power spectra from simulations. The power spectra are very well recovered within one standard error.

\begin{figure}
    \centering
    \includegraphics[width=\linewidth]{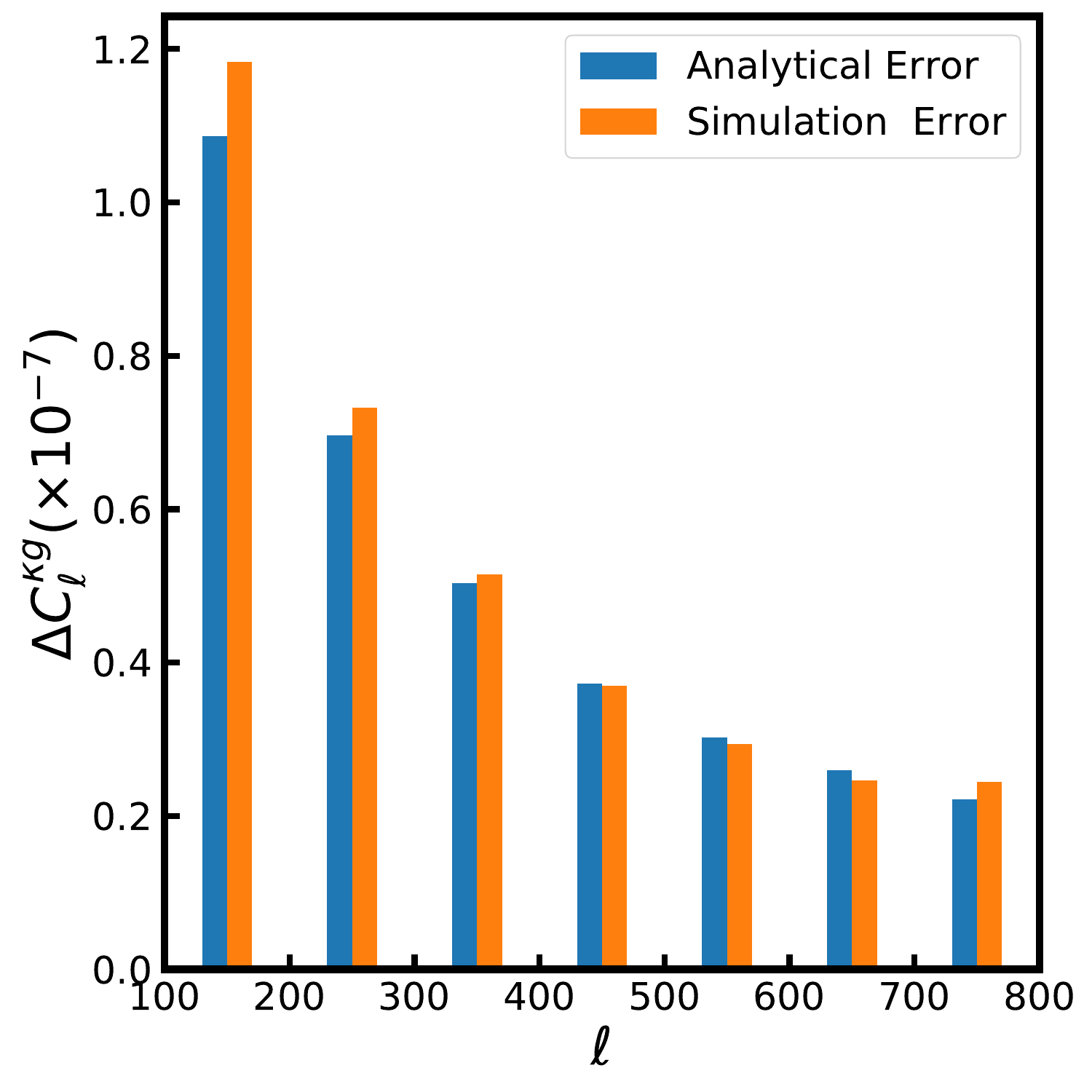}
    \caption{Comparison of simulation (orange) and analytical (blue) errors for cross-power spectrum of HS-82 field. The orange bars are computed from the square root of the diagonal of Eq. \ref{eq:cov_simul}. The blue bars are calculated from Eq. \ref{eq:error_covariance}}
    \label{fig:kg_sim_err_comp}
\end{figure}

In Fig.~\ref{fig:kg_sim_err_comp} we also show the comparison of errors on cross-power spectrum estimated for simulations from the square root of the diagonal of Eq.~\ref{eq:cov_simul} and errors estimated analytically from Eq. \ref{eq:error_covariance}.

\begin{figure}
    \centering
    \includegraphics[width=\linewidth]{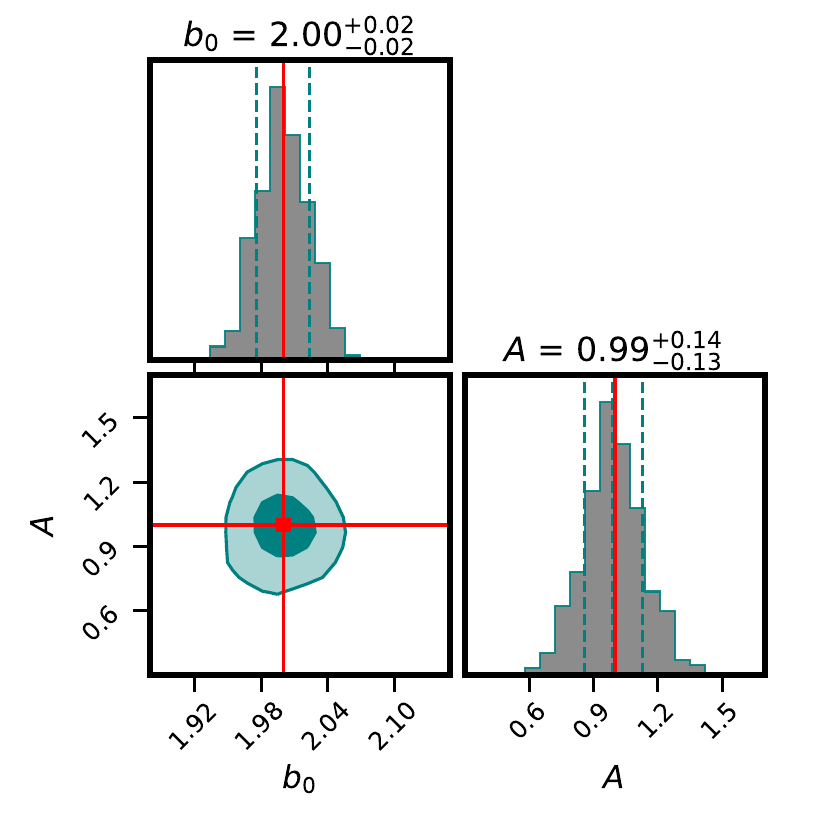}
    \caption{Distribution of estimated parameters $b_{0}$ and $A$ for 500 simulations of HS-82 field. The lighter and darker contours represent the $68\%$ and $95\%$ confidence levels. The three vertical lines are the median value of marginalised posteriors and $\pm 1\sigma$ errors. The red lines show the true values of parameter $b=2$ and $A=1$ used for simulations.}
    \label{fig:dist_params_500_sim_all_objects_alpha1}
\end{figure}

To check whether estimation of the parameters is unbiased in Fig.~\ref{fig:dist_params_500_sim_all_objects_alpha1} we show the comparison of true values of the parameters that we use in simulations, i.e.~$b_{0}=2$ and $A=1$ (red lines), with the distribution of the parameters estimated from 500 simulations of HS-82 field with the procedure mentioned in section \ref{sec:likeli}. As we can see the distribution of parameters is very well centered on the true values used for simulations which prove that our algorithm provides an unbiased estimation of the parameters. We find similar results for other fields as well.

\section{Results}\label{sec:results}
We now present the results of extracting power spectrum and galaxy linear bias parameter $b_{0}$ and amplitude of cross-power spectrum $A$ using the procedures described in sections \ref{sec:methodology} and section \ref{sec:likeli}.
\subsection{Power Spectra}

\begin{figure*}
    \begin{subfigure}[b]{0.25\linewidth}
        \centering
        \includegraphics[width=\linewidth]{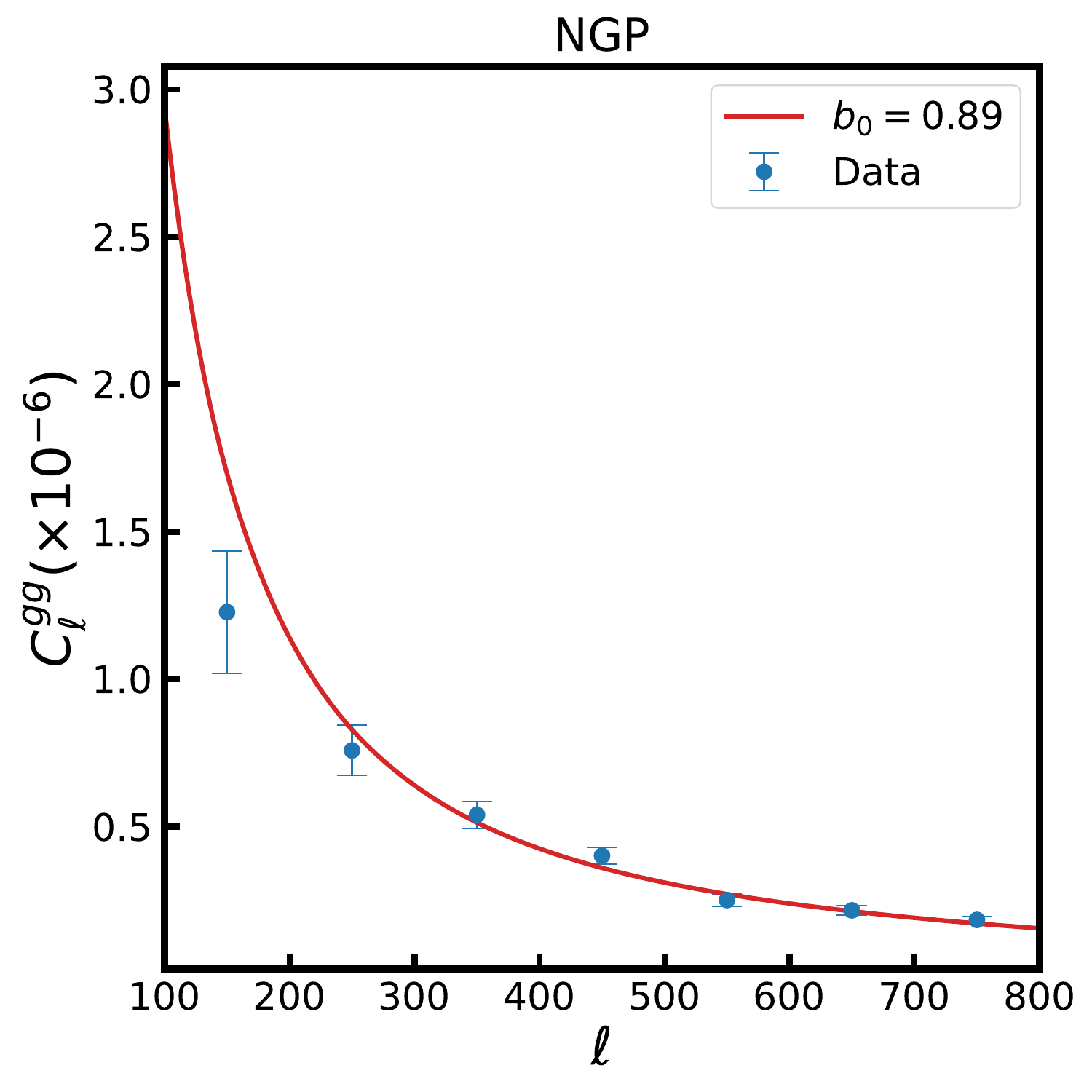}
    \end{subfigure}%
    \begin{subfigure}[b]{0.25\linewidth}
        \centering
        \includegraphics[width=\linewidth]{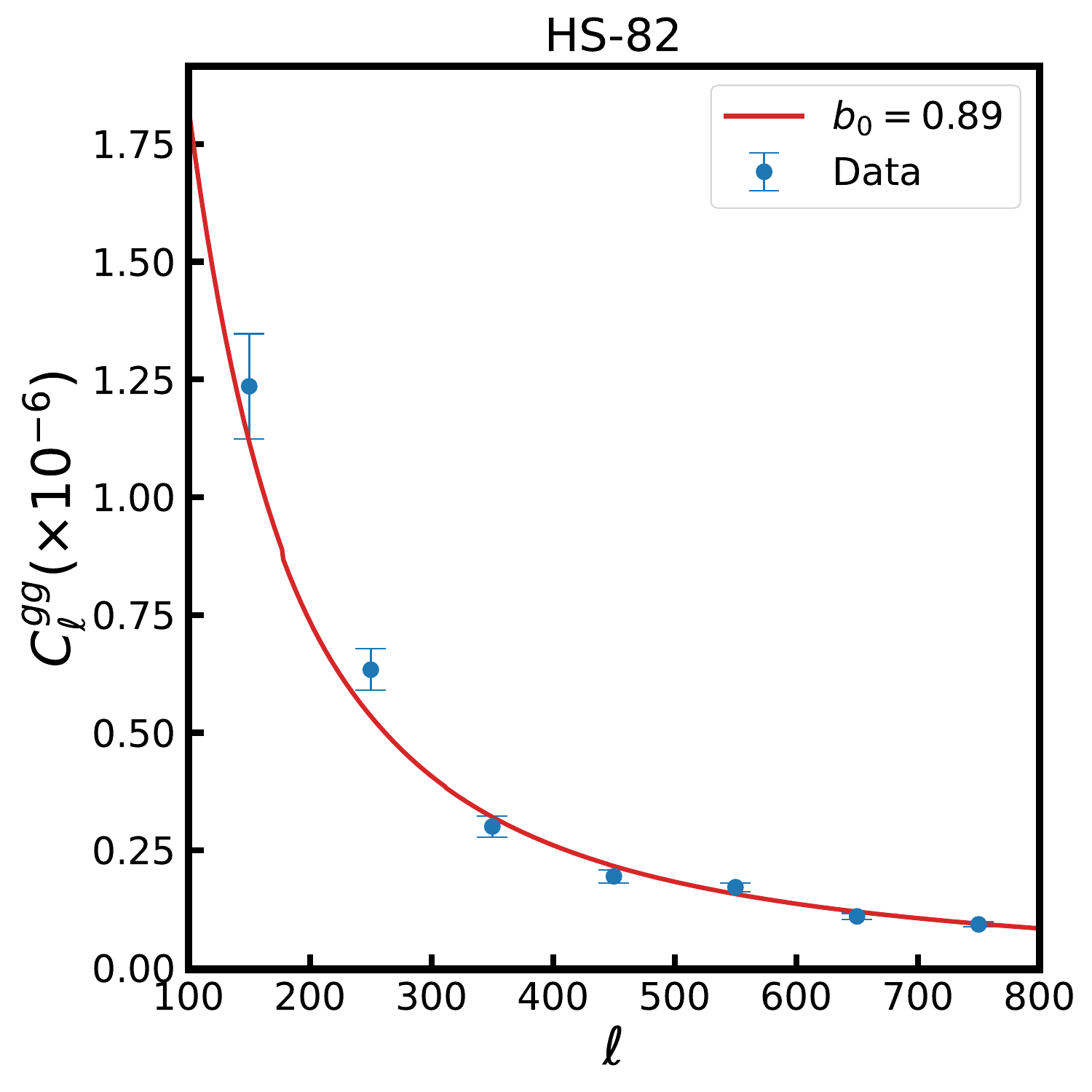}
    \end{subfigure}%
    \begin{subfigure}[b]{0.25\linewidth}
        \centering
        \includegraphics[width=\linewidth]{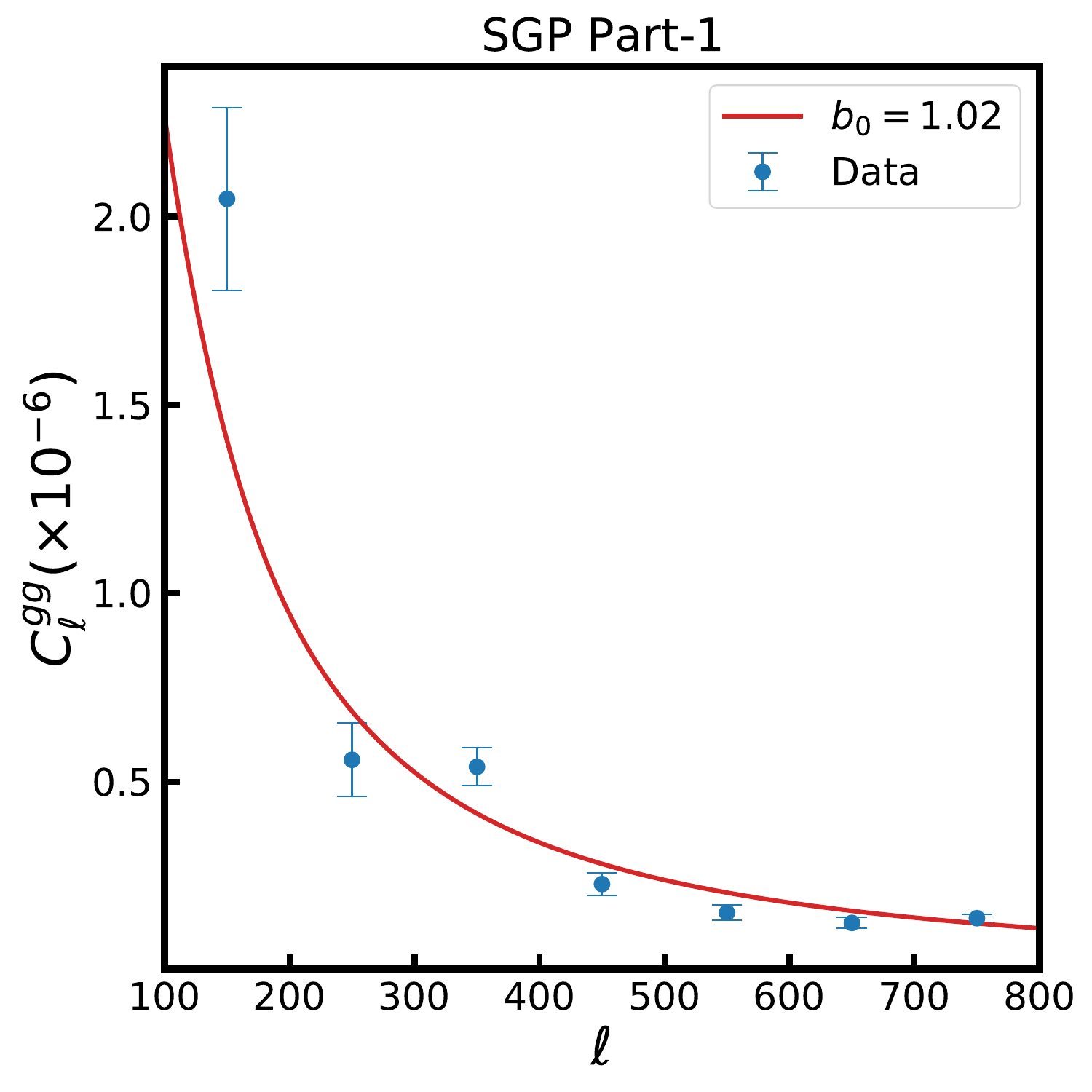}
    \end{subfigure}%
    \begin{subfigure}[b]{0.25\linewidth}
        \centering
        \includegraphics[width=\linewidth]{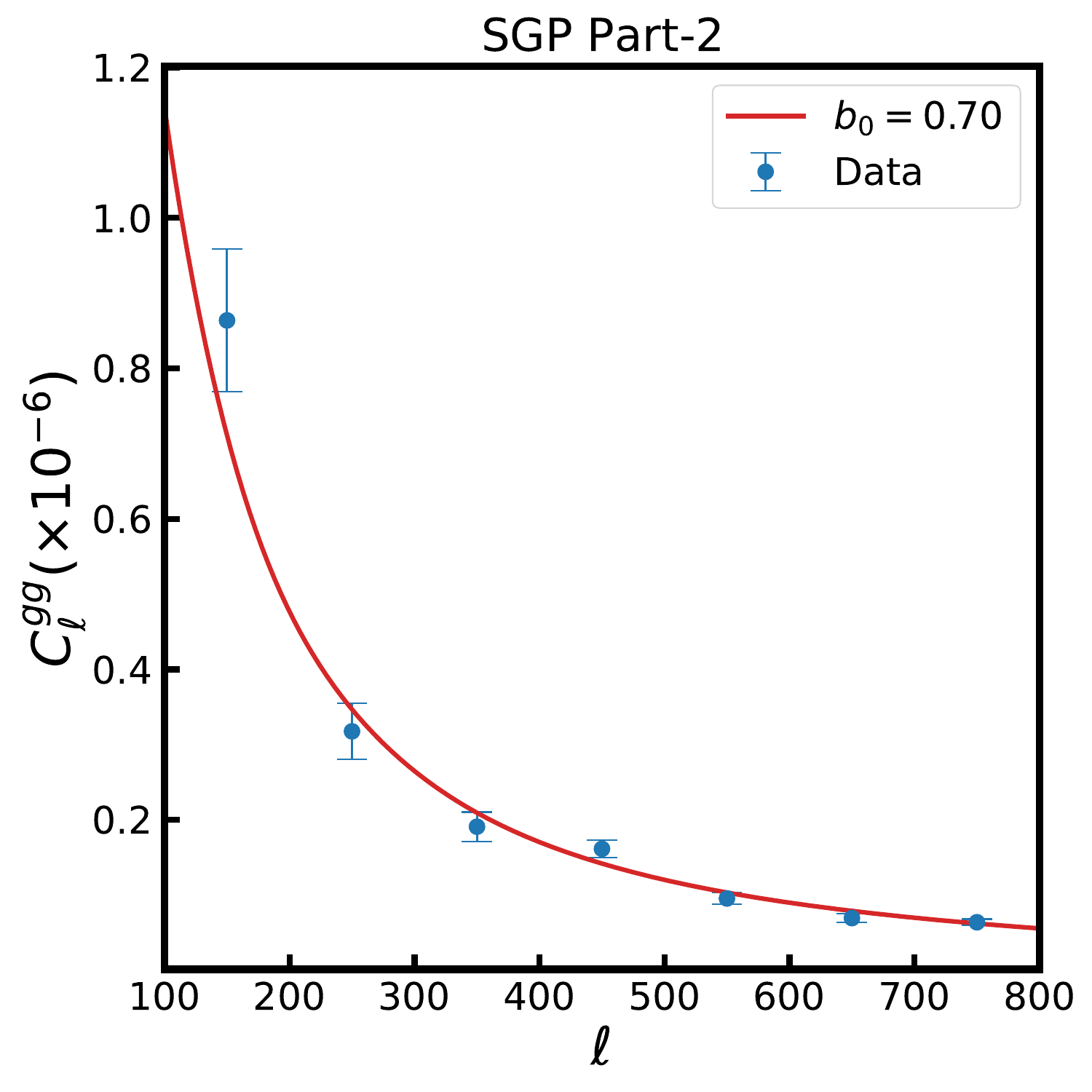}
    \end{subfigure}\\[5ex]
    \begin{subfigure}[b]{0.25\linewidth}
        \centering
        \includegraphics[width=\linewidth]{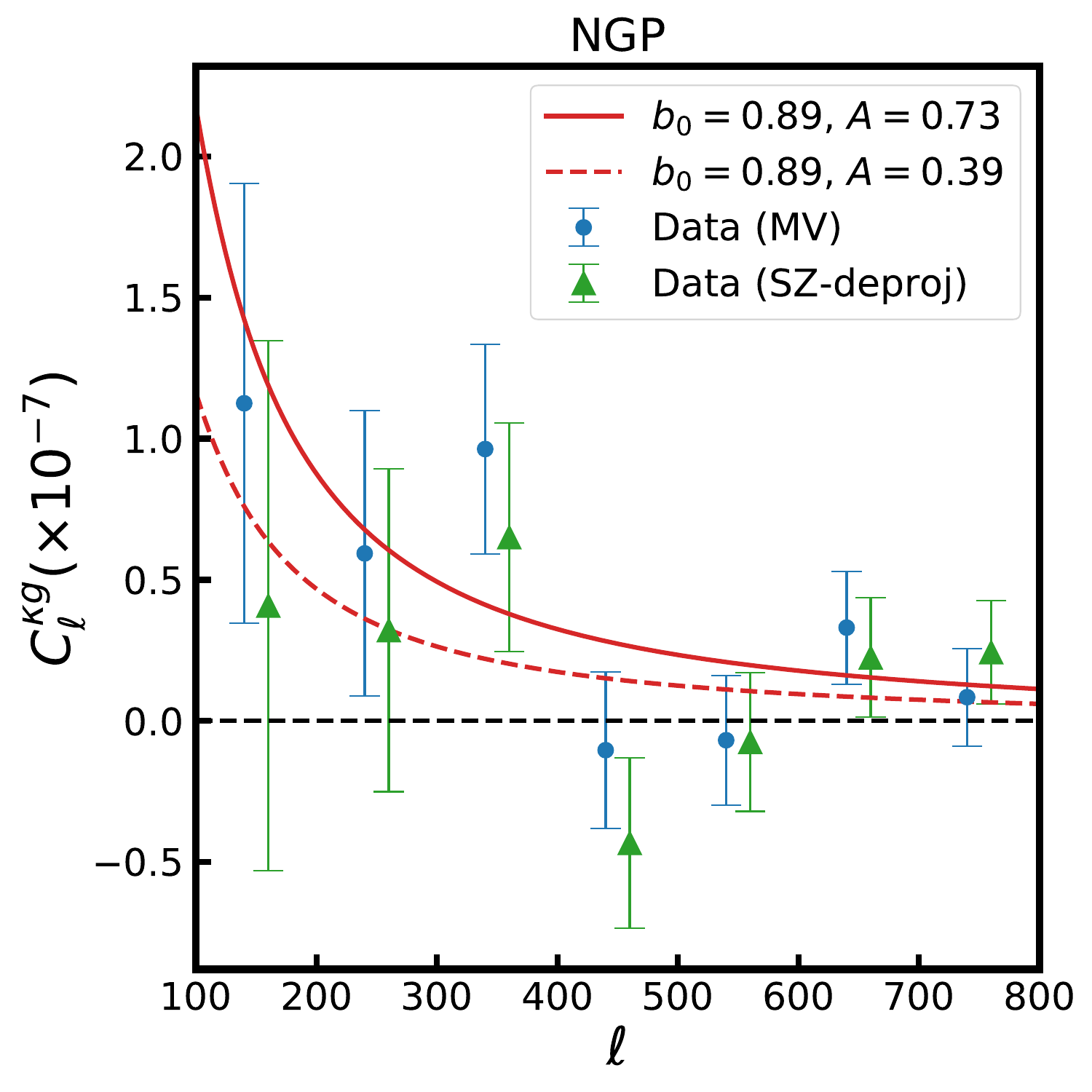}
    \end{subfigure}%
     \begin{subfigure}[b]{0.25\linewidth}
        \centering
        \includegraphics[width=\linewidth]{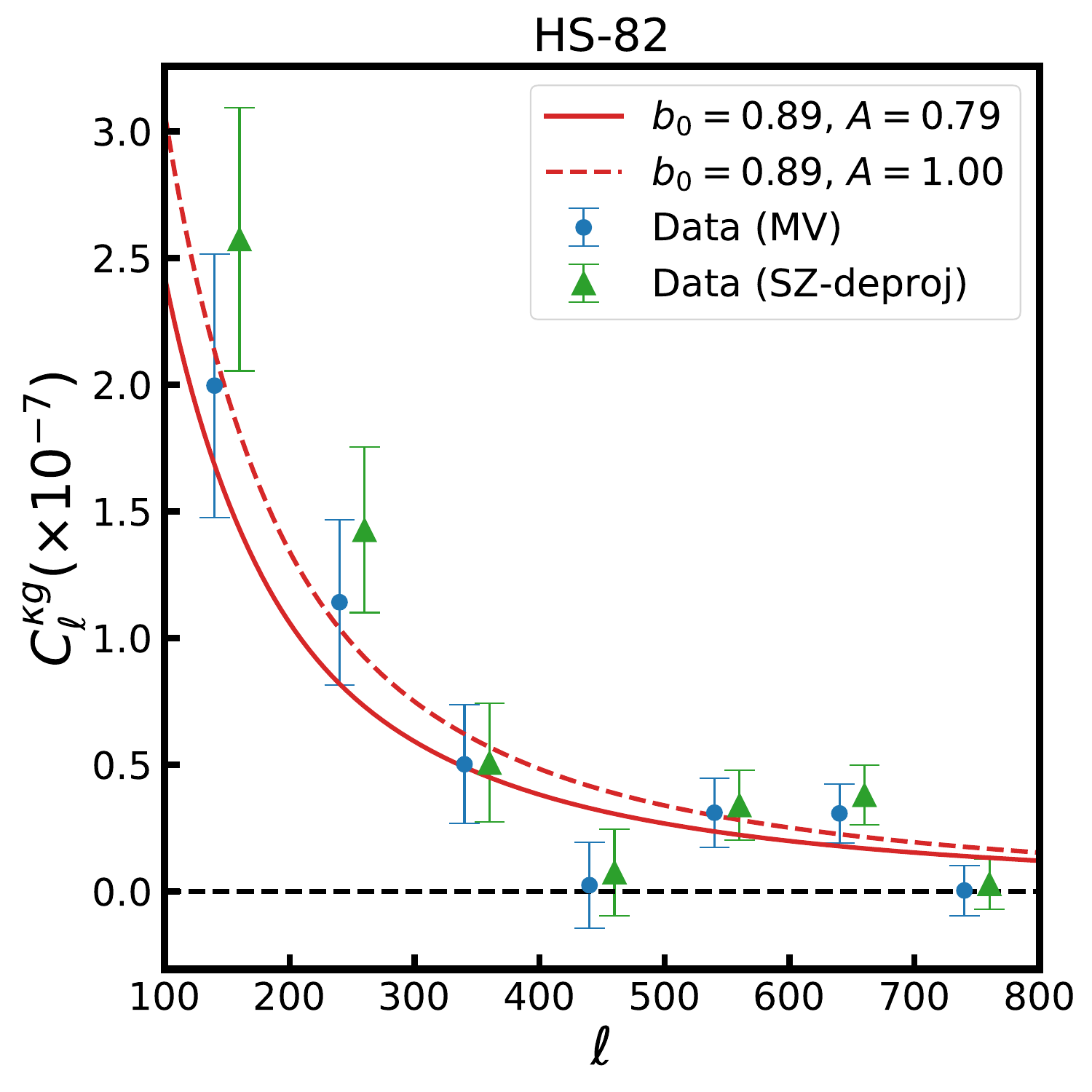}
    \end{subfigure}%
     \begin{subfigure}[b]{0.25\linewidth}
        \centering
        \includegraphics[width=\linewidth]{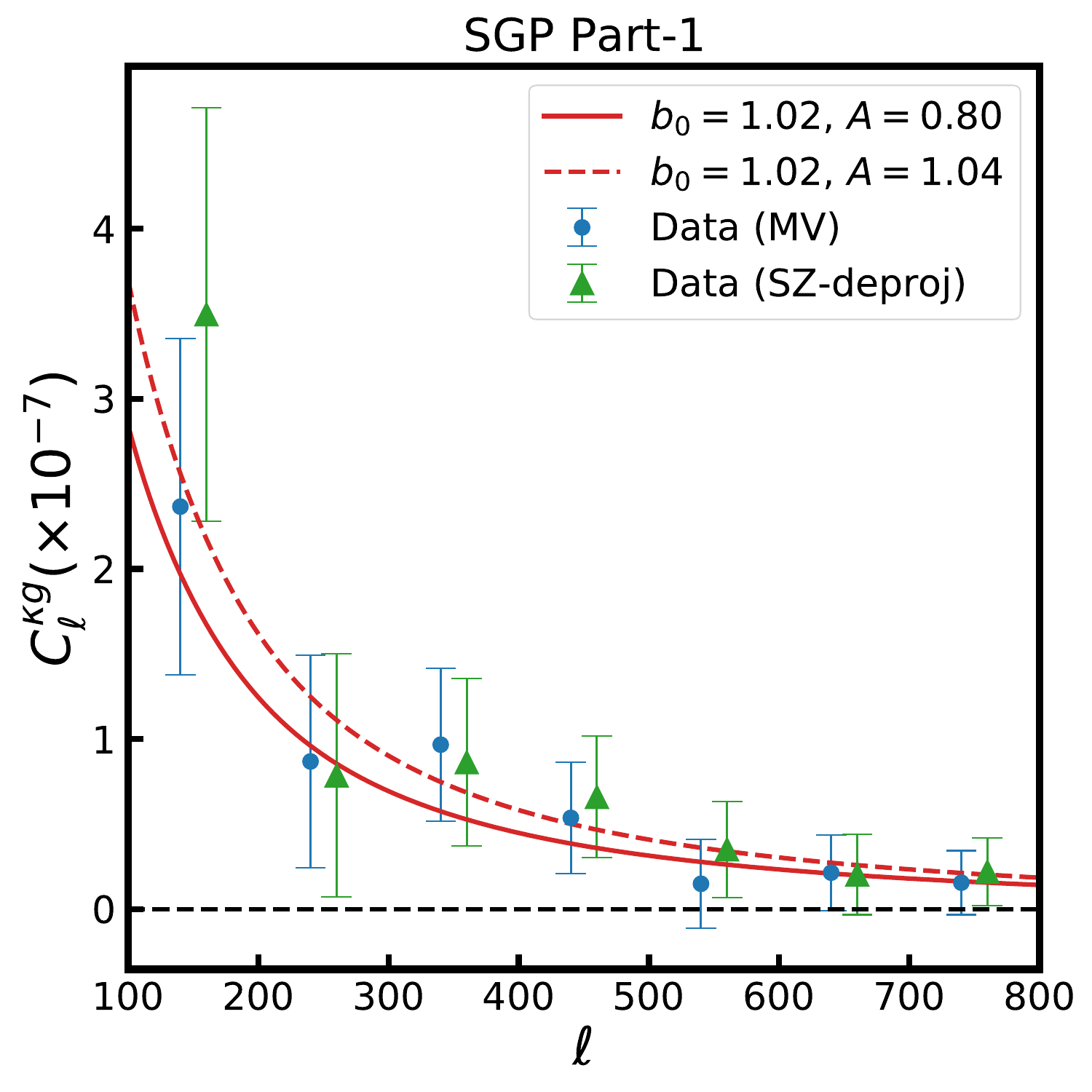}
    \end{subfigure}%
     \begin{subfigure}[b]{0.25\linewidth}
        \centering
        \includegraphics[width=\linewidth]{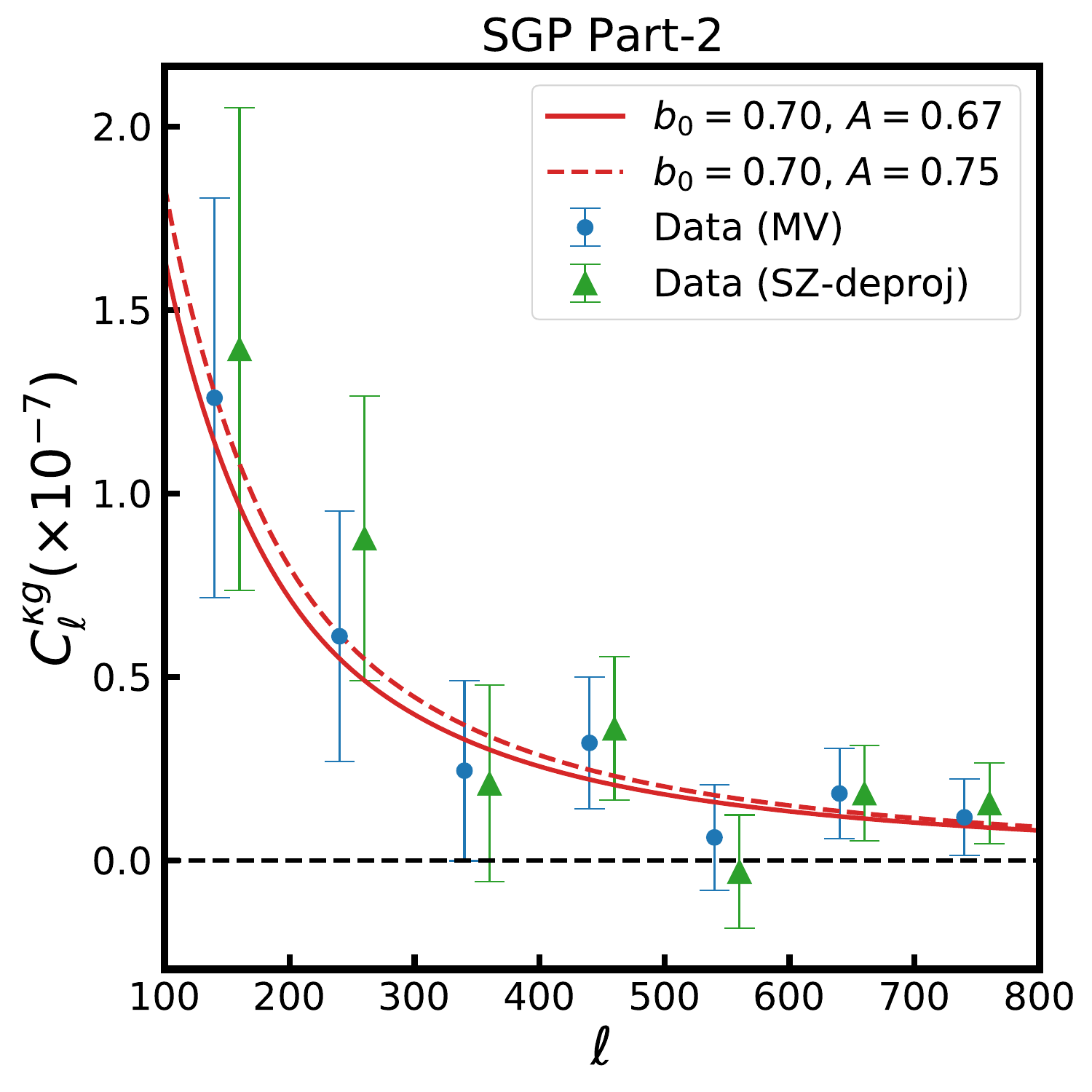}
    \end{subfigure}
    \caption{Galaxy auto-power spectra (\textit{Top}) and cross-power spectra (\textit{Bottom}) for HELP fields. The blue circles represent the measurements of cross-correlation signal using MV map while green triangles are for SZ-deproj map. The red solid line is the theoretical power spectrum computed using the best-fit values obtained from likelihood analysis using MV map and the red dashed line is the theoretical line for SZ-deproj map.}
    \label{fig:power_spectra_all_original}
\end{figure*}

The noise subtracted galaxy auto-power spectra and cross-power spectra for all four HELP fields are shown in Fig. \ref{fig:power_spectra_all_original}. The errors are estimated from the diagonal of the analytical covariance matrix described in section \ref{sec:errors} using the best fit values of $b_{0}$ and $A$.

Before going over parameter estimation, we discuss the results of the null hypothesis. We calculate the probability that the estimated signal is consistent with no correlation between galaxy over-density and lensing convergence fields. The $\chi^{2}_{null}$ values are calculated using

\begin{equation}
    \chi^{2}_{null} = \hat{C}_{L}^{\kappa g}(Cov_{LL'}^{\kappa g})^{-1}\hat{C}_{L'}^{\kappa g}
    \label{eq:null_chi_square}
\end{equation}
The $\chi^{2}_{null}$ values for cross-power spectra from all four patches and their associated $p$-values are presented in Table \ref{tab:null_hypothesis_original} with $\nu = 7$ being the number of degrees of freedom.

\begin{table}
	\centering
	\caption{Result of no correlation hypothesis rejection for HELP patches using MV and SZ-deproj convergence maps.}
	\label{tab:null_hypothesis_original}
	\begin{tabular}{lcc||cc} 
		\hline\hline
		Patch & \multicolumn{2}{c}{MV} & \multicolumn{2}{c}{SZ-deproj}\\
		\cline{2-5}
	          & & & &\\
		      & $\chi^{2}_{null}$/$\nu$ & $p$-value & $\chi^{2}_{null}$/$\nu$ & $p$-value \\
		      & & & &\\
		\hline
		NGP & $13.4/7$ & $6.32\times 10^{-2}$ & $8.14/7$ & $0.32$\\
		& & & &\\
 		HS-82 & $41.5/7$ & $6.57\times 10^{-7}$ & $54.9/7$ & $1.59\times 10^{-9}$\\
 		& & & &\\
 		SGP Part 1 & $16.9/7$ & $1.80\times 10^{-2}$ & $19.5/7$ & $6.69\times 10^{-3}$\\
 		& & & &\\
 		SGP Part 2 & $16.4/7$ & $2.16\times 10^{-2}$ & $17.7/7$ & $1.36\times 10^{-2}$\\
		\hline
	\end{tabular}
\end{table}

\begin{figure}
    \centering
    \includegraphics[width=\linewidth]{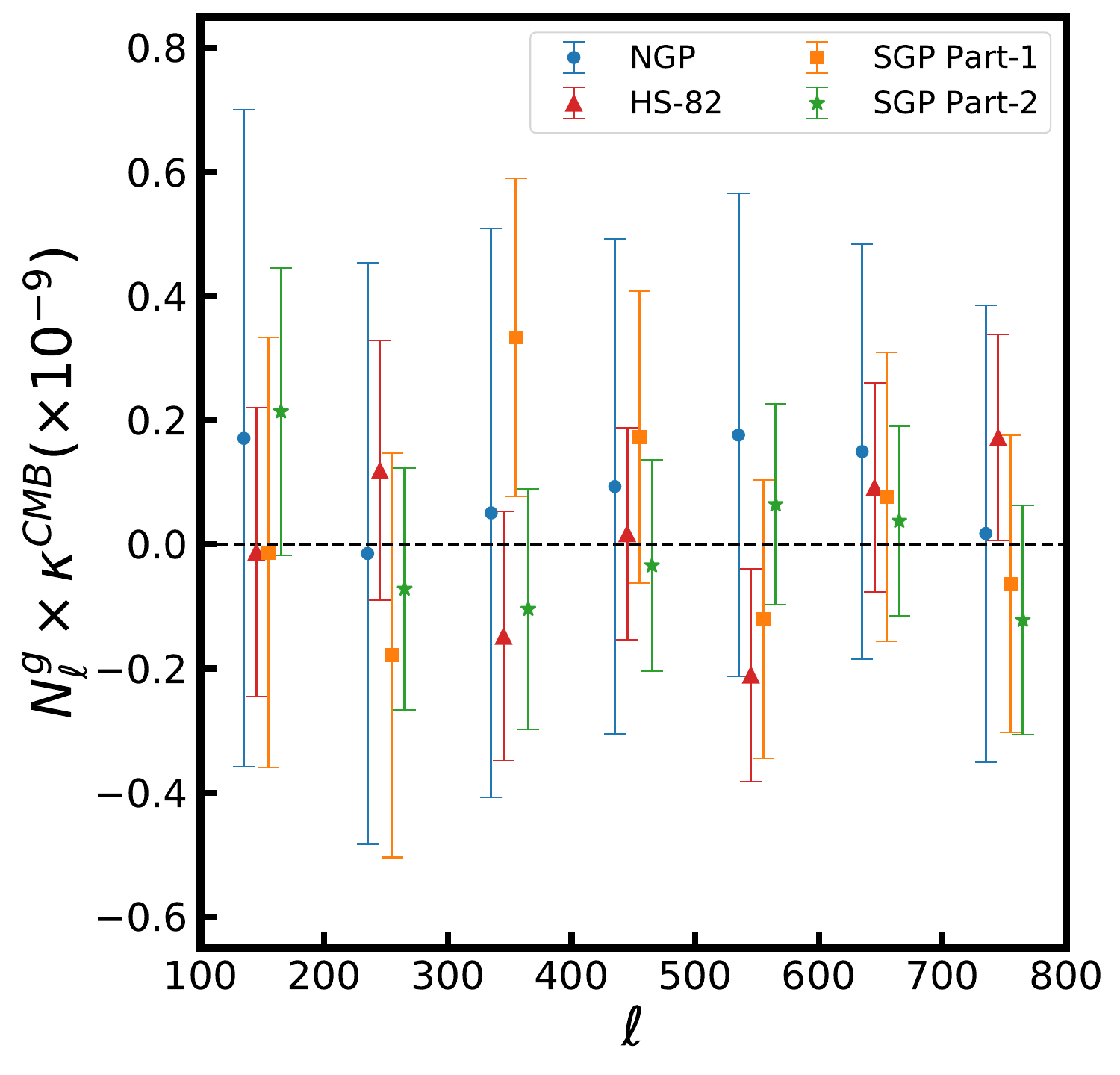}
    \caption{Mean correlation between galaxy shot noise obtained from jackknifing approach and CMB lensing convergence map. No significant signal is detected for any patch.}
    \label{fig:noise_correlation}
\end{figure}

Another null test we perform is cross-correlating galaxy shot noise, estimated by a jackknifing approach (\citeauthor{Ando2018} \citeyear{Ando2018}; \citeauthor{Bianchini2018} \citeyear{Bianchini2018}) with true \textit{Planck} CMB convergence map. We randomly split the galaxy catalogue into two over-density maps $\delta_{g}^{1}$ and $\delta_{g}^{2}$. Then, the map $(\delta_{g}^{1}+\delta_{g}^{2})/2$ contains both signal and noise, while $(\delta_{g}^{1}-\delta_{g}^{2})/2$ is noise-only map. Fig. \ref{fig:noise_correlation} shows the mean correlation of 500 galaxy shot noise obtained from jackknifing method for all patches with lensing convergence map. The error bars are computed from the covariance matrices obtained from these 500 correlations. No significant cross-correlation signal is detected for any patch and we obtain $\chi^{2}/\nu=0.083, 0.537, 0.429 \text{ and } 0.283$ with $\nu=7$ degrees of freedom, with probability of no correlation $p = 0.99,0.81,0.88$ and $0.96$ for NGP, HS-82, SGP Part-1 and SGP Part-2, respectively.

\subsection{ Galaxy linear bias parameter and Amplitude}\label{subsec:parameters}

Estimated values of parameters for real data using MV lensing map are presented in Table \ref{tab:likeli_result_original_kg} and those using SZ-deproj lensing map are presented in Table \ref{tab:likeli_result_all_objects_same_lmin_sz}. Constraints on galaxy linear bias parameter and amplitude of cross-power spectrum were obtained by sampling the likelihood function given in Eq.~\ref{eq:joint_likeli} for the joint power spectrum $\hat{C}_{L}$. We also compute these parameters by independently sampling the likelihood function for measured galaxy auto-power spectrum, $\hat{C}_{L}^{gg}$ and cross-power spectrum, $\hat{C}_{L}^{\kappa g}$ only.

To give some idea on goodness of fit of theoretical power spectra to observations in the tables we provide also $\chi^{2}$ and $p$-values. We evaluate the $\chi^{2}$ values using the expression: $\chi^{2}=[\hat{C}_{L}^{\kappa g}-A^{bf}C_{L}^{\kappa g}(b^{bf}_{0})](Cov_{LL'}^{\kappa g})^{-1}[\hat{C}_{L'}^{\kappa g}-A^{bf}C_{L'}^{\kappa g}(b^{bf}_{0})]$, where $b^{bf}_{0}$ and $A^{bf}$ stand for best-fit values of galaxy linear bias parameter $b_{0}$ and amplitude of cross-power spectrum $A$. The $p$-values are calculated using the standard theoretical $\chi^{2}$ distribution function with $\nu=5$ degrees of freedom. Using MV map, we obtain a value of $\chi^{2}/\nu \sim 0.3$ or SGP Part-1 and Part-2, $\sim 1.3$ for NGP and $\sim 1.6$ for HS-82 field. The SZ-deproj gives similar $\chi^{2}/\nu$ values of $\sim 1.4$ for NGP and $\sim 0.3$ for SGP Part-1, while larger $\chi^{2}/\nu$ values of $\sim 0.66$ for SGP Part-2 and $\sim 2.1$ for HS-82.

\begin{table*}
	\centering
	\caption{Galaxy linear bias parameter and cross-correlation amplitude for HELP patches with MV lensing potential map using both separate and joint likelihood functions.}
	\label{tab:likeli_result_original_kg}
	\begin{tabular}{lccccccc} 
		\hline\hline
		Patch & $gg$ & \multicolumn{2}{c}{$\kappa g$} & \multicolumn{2}{c}{$\kappa g, gg$} & $\chi^{2}$/$\nu$ & $p$-value\\
		\cline{3-6}
		 & & $b_{0}$ & $A$ & $b_{0}$ & $A$ & & \\
		\hline
		NGP &  $0.89_{-0.01}^{+0.01}$ & $0.97_{-0.26}^{+0.42}$ & $0.61_{-0.30}^{+0.32}$ & $0.89_{-0.01}^{+0.01}$ & $0.73_{-0.24}^{+0.24}$ & 6.5/5 & 0.262\\
		\vspace{1mm}\\
		HS-82 & $0.89_{-0.01}^{+0.01}$ & $1.08_{-0.27}^{+0.45}$ & $0.71_{-0.25}^{+0.28}$ & $0.89_{-0.01}^{+0.01}$ & $0.79_{-0.14}^{+0.14}$ & 8.0/5 & 0.155\\
		\vspace{1mm}\\
		SGP Part-1 & $1.02_{-0.02}^{+0.02}$ & $0.45_{-0.13}^{+0.20}$ & $2.16_{-0.71}^{+0.89}$ & $1.02_{-0.02}^{+0.02}$ & $0.80_{-0.23}^{+0.23}$ & 1.4/5 & 0.919\\
		\vspace{1mm}\\
		SGP Part-2 & $0.70_{-0.01}^{+0.01}$ & $0.26_{-0.10}^{+0.14}$ & $1.97_{-0.72}^{+1.15}$ & $0.70_{-0.01}^{+0.01}$ & $0.67_{-0.18}^{+0.18}$ & 1.3/5 & 0.938\\
		\hline
	\end{tabular}
\end{table*}

\begin{figure*}
    \begin{subfigure}[b]{0.25\linewidth}
        \centering
        \includegraphics[width=\linewidth]{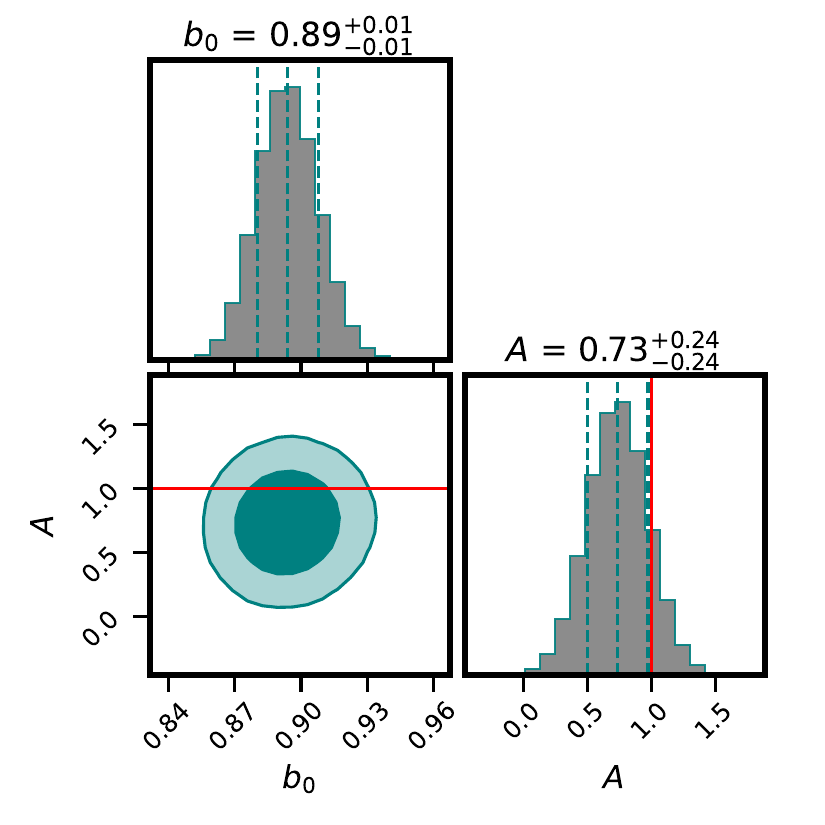}
        \caption{NGP}
    \end{subfigure}%
    \begin{subfigure}[b]{0.25\linewidth}
        \centering
        \includegraphics[width=\linewidth]{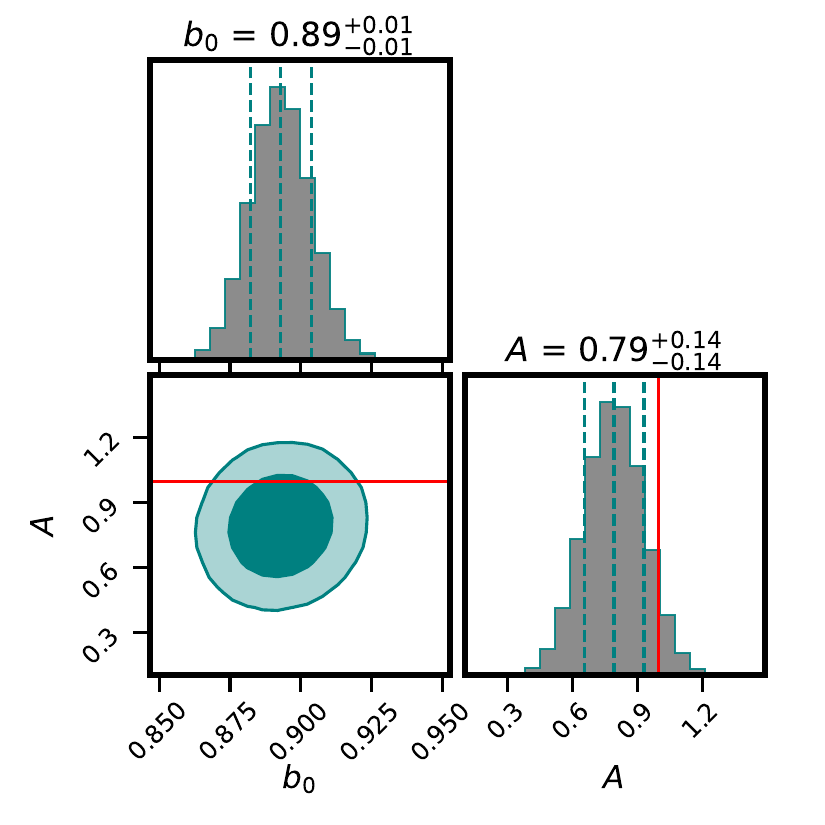}
        \caption{HS-82}
    \end{subfigure}%
    \begin{subfigure}[b]{0.25\linewidth}
        \centering
        \includegraphics[width=\linewidth]{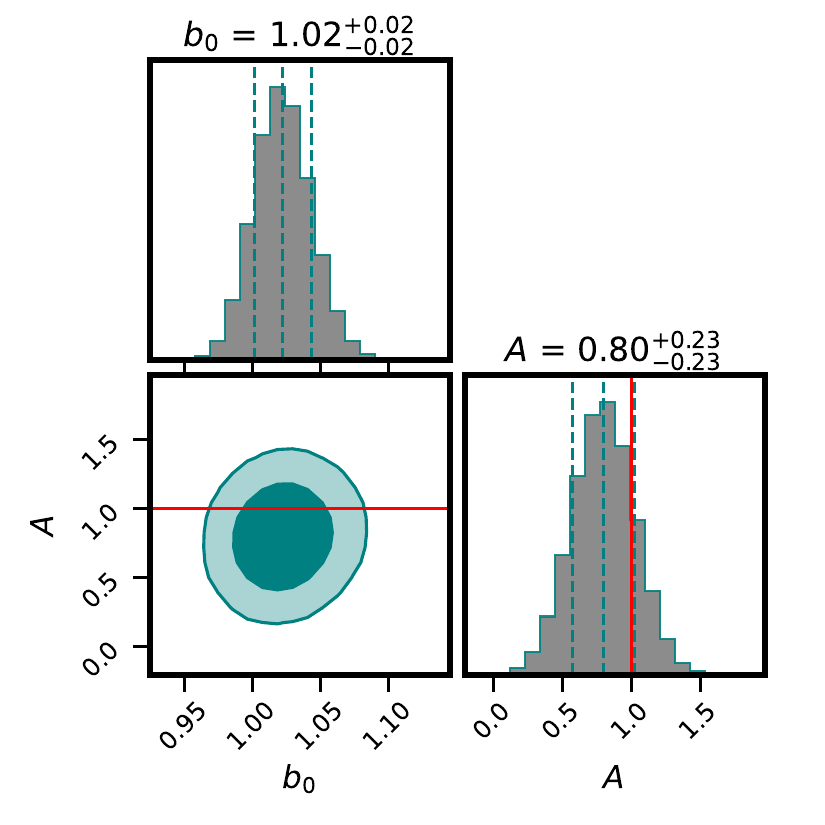}
        \caption{SGP Part-1}
    \end{subfigure}%
    \begin{subfigure}[b]{0.25\linewidth}
        \centering
        \includegraphics[width=\linewidth]{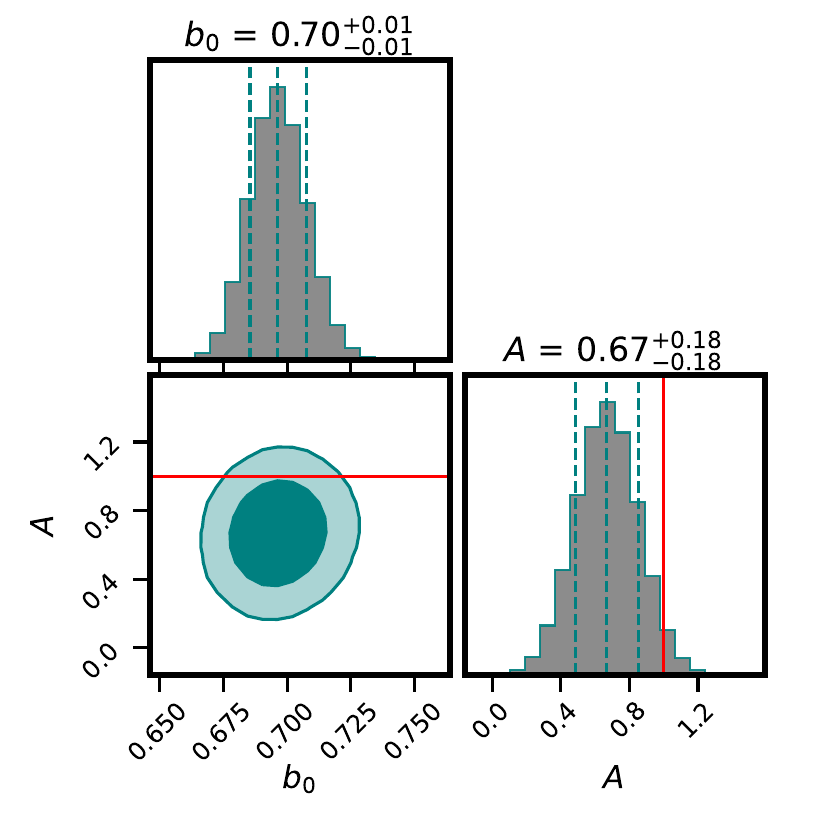}
        \caption{SGP Part-2}
    \end{subfigure}
    \caption{Posteriors of parameter obtained from Maximum Likelihood Estimation for all HELP patches with $68\%$ and $95\%$ confidence contours shown in darker and lighter shades, respectively. The three vertical lines are the median value of posterior and $\pm 1\sigma$ errors. The red line represents the value of $A=1$ for standard $\Lambda$CDM.}
    \label{fig:results_mle_original}
\end{figure*}

\begin{table*}
	\centering
	\caption{Galaxy linear bias parameter and cross-correlation amplitude for HELP patches with SZ-deproj lensing potential map using both separate and joint likelihood functions.}
	\label{tab:likeli_result_all_objects_same_lmin_sz}
	\begin{tabular}{lccccccc} 
		\hline\hline
		Patch & $gg$ & \multicolumn{2}{c}{$\kappa g$} & \multicolumn{2}{c}{$\kappa g, gg$} & $\chi^{2}$/$\nu$ & $p$-value\\
		\cline{3-6}
		 & & $b_{0}$ & $A$ & $b_{0}$ & $A$ & & \\
		\hline
		NGP &  $0.89_{-0.01}^{+0.01}$ & $1.00_{-0.27}^{+0.43}$ & $0.29_{-0.28}^{+0.29}$ & $0.89_{-0.01}^{+0.01}$ & $0.39_{-0.27}^{+0.27}$ & 6.9/5 & 0.229\\
		\vspace{1mm}\\
		HS-82 & $0.89_{-0.01}^{+0.01}$ & $1.26_{-0.32}^{+0.51}$ & $0.75_{-0.26}^{+0.29}$ & $0.89_{-0.01}^{+0.01}$ & $1.00_{-0.14}^{+0.14}$ & 10.5/5 & 0.062\\
		\vspace{1mm}\\
		SGP Part-1 & $1.02_{-0.02}^{+0.02}$ & $0.50_{-0.13}^{+0.21}$ & $2.34_{-0.75}^{+0.86}$ & $1.02_{-0.02}^{+0.02}$ & $1.04_{-0.26}^{+0.26}$ & 1.3/5 & 0.938\\
		\vspace{1mm}\\
		SGP Part-2 & $0.70_{-0.01}^{+0.01}$ & $0.53_{-0.14}^{+0.23}$ & $1.08_{-0.39}^{+0.45}$ & $0.70_{-0.01}^{+0.01}$ & $0.75_{-0.21}^{+0.21}$ & 3.3/5 & 0.645\\
		\hline
	\end{tabular}
\end{table*}
In Fig.~\ref{fig:results_mle_original}, the 2-dimensional posterior distributions in the $(b_{0},A)$ plane for cross-correlation of HELP fields with MV map and corresponding marginalised distributions for each parameter are shown. For the joint analysis of the MV convergence map, we find agreement between estimated cross-correlation amplitude $A$ and the expected value within $\sim 1\,\sigma$ for NGP and SGP Part-1, within $\sim 1.5\,\sigma$ for HS-82 and within $\sim 2\,\sigma$ for SGP Part-2. For the SZ-deproj map, the amplitudes are larger and in agreement with one within $\sim 1\,\sigma$ for all fields, except the NGP field for which the estimated amplitude takes a value of 0.39 and the significance of the discrepancy is 2.2\,$\sigma$. The disparities between results for MV and SZ-deproj lensing maps can be caused by differences in the maps, in particular using CMB polarisation data for MV map and removing SZ signal from the CMB temperature map for SZ-deproj lensing map.

The estimated galaxy linear bias parameter for the joint analysis takes the same values as the ones estimated from the galaxy auto-power spectrum. It shows that the parameter is in principle entirely constrained by the latter. The parameter takes value in a range from 0.70 for SGP Part-2 to 1.02 for SGP Part-1 field with less than $2\,\%$ errors. In the case of NGP an HS-82 fields the parameter takes the same value of 0.89. Differences between values of the bias parameter can be related with some variation in selection of objects for different fields as not all of them are observed by the same set of galaxy surveys. Table \ref{apndx_tab:bandwise_coverage_help_fields} shows the percentage of objects observed by different surveys for a given HELP field used in this study. As it was already mentioned in Sect.~\ref{sec:gal_data}, SGP Part-1 field is covered by KiDS survey while SGP Part-2 field by DES survey. On the other hand, contrary to SGP fields, a large fraction of objects in NGP and HS-82 fields are observed by the PanSTARRS survey what can result in selection of similar objects in both fields and the same value of the bias parameter.

\section{Discussion and Conclusions}\label{sec:discuss}

The work presented in this paper is the first estimation of galaxy linear bias parameter $b_{0}$ and amplitude of cross-power spectrum $A$ for the HELP catalogue. The correlation between MV CMB lensing potential and the distribution of galaxies is found to be lower than expected, but within 2\,$\sigma$ for all patches. We now discuss possible reasons that can lead to a smaller value of the amplitude of cross-power spectrum $A$.

As is shown in Fig.~\ref{fig:dist_params_500_sim_all_objects_alpha1} and discussed in section \ref{subsec:parameters}, we do not find any systematics in the estimation of the parameters from simulations, so we can exclude as a possible reason for observed deviation some bias introduced by algorithms used for data analysis. As a consistency check, we also applied our algorithms to the dataset used in studies presented in \cite{Bianchini2015}. We reproduced the galaxy linear bias parameter and amplitude of cross-spectrum precisely within $1\,\sigma$ values as quoted in table 3 of \cite{Bianchini2015}. This further solidifies the procedure we use to estimate parameters in this work.

\subsection{CIB contamination} \label{sec:cib}

One possible reason for a smaller value of the amplitude can be the correlation of galaxies with residuals of Cosmic Infrared Background (CIB) emission which leaked into the lensing map through the CMB temperature maps used for the lensing estimation. As we can see in Fig.~23 from \cite{Planck2020VIII}, CIB-induced bias for the MV and SZ-deproj CMB lensing map auto-power spectrum is at the sub-percent level for the range of multipoles used in our analysis. Such a small CIB contribution is a result of using the SMICA CMB map for the lensing map estimation in the 2018 \textit{Planck} data release, which significantly reduces CIB contamination. Nevertheless, it can be larger for the cross-power spectrum considered in this work. We did not estimate CIB contribution to cross-correlation leaving it for future studies, however, it is worth noticing that the redshift range of the HELP galaxies is rather low comparing to the redshifts of CIB sources, so we do not expect substantial cross-correlation between them. Because CIB-induced bias for the SZ-deproj lensing map is roughly two times larger than for the MV lensing map, we decided to use the latter in our baseline analysis.

\subsection{Magnification Bias}

\begin{figure}
    \centering
    \includegraphics[width=\linewidth]{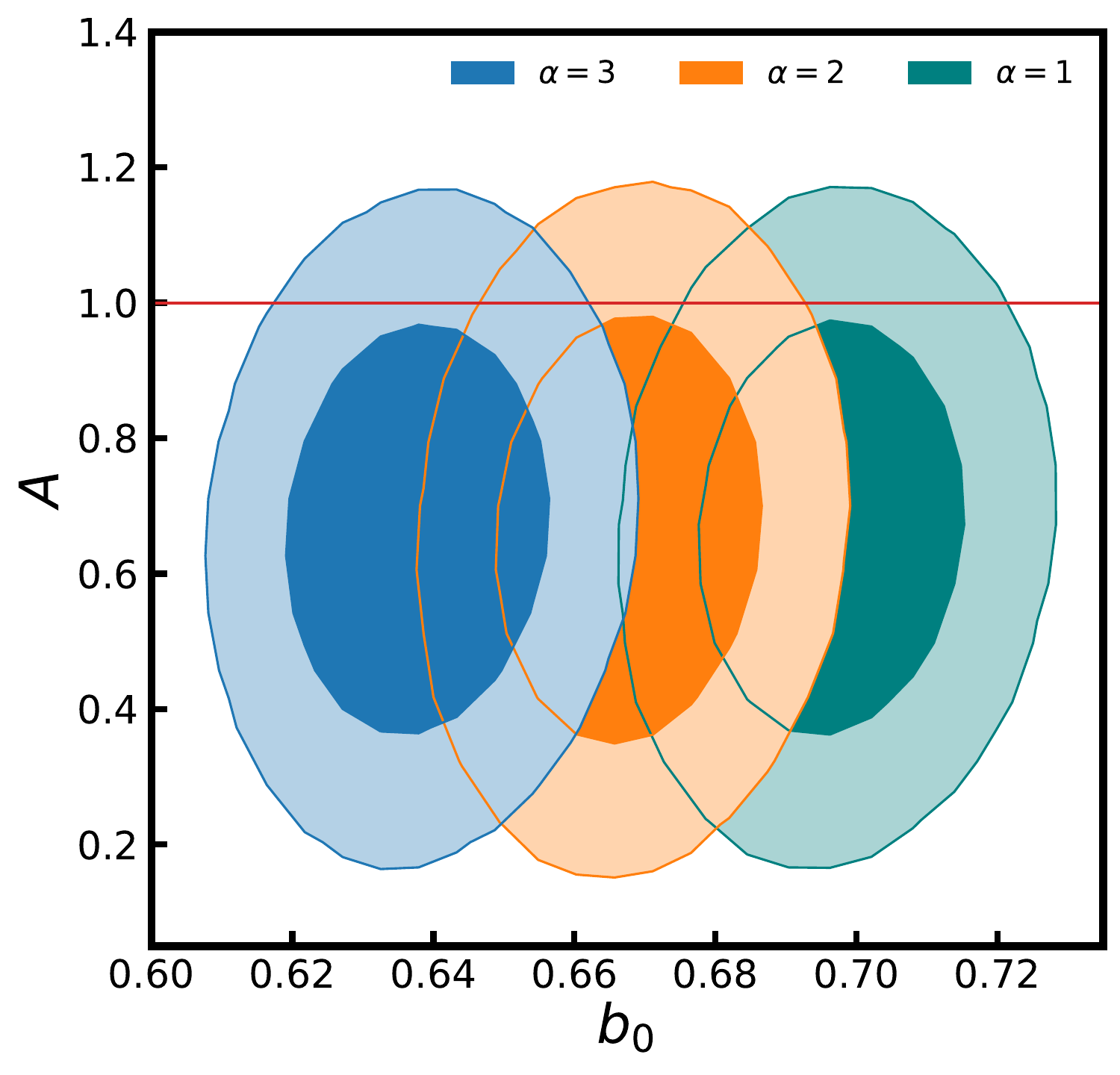}
    \caption{Effect of $\alpha$ on the estimated values of cross-correlation amplitude $A$ and galaxy linear bias parameter $b_{0}$ for SGP Part-2. There is no significant effect on $A$.}
    \label{fig:testing_mag_bias}
\end{figure}

The second term in Eq.~\ref{eq:galaxy_kernel} accounts for the modification of the observed density of background sources due to weak gravitational lensing by foreground objects. We measured the value of $\alpha = 1$ for all HELP patches used in our study by fitting a straight line to ${\log {N(>S)}}$ distribution. In this section we look at the change in parameters, $b_{0}$ and $A$, for $\alpha = 2$ and $\alpha=3$. Fig.~\ref{fig:testing_mag_bias} presents the 2-dimensional posterior contours in the $(b_{0},A)$ plane showing change in parameters measured from SGP Part-2 for different values of $\alpha$. From the joint analysis of galaxy auto-power spectrum and cross-power spectrum we get $b_{0}=0.67\pm 0.01$ and $A=0.67\pm 0.18$ for $\alpha = 2$ and $b_{0}=0.64\pm 0.01$ and $A=0.67\pm 0.18$ for $\alpha = 3$. Higher values of $\alpha$ lead to smaller values of galaxy linear bias parameter $b_{0}$ but do not affect the amplitude $A$. We see a similar trend also in other patches. For relatively shallow surveys such as HELP, cross-correlation with the convergence field is weak. Hence, we do not see significant changes in the cross-correlation amplitude because, for different values of $\alpha$, changes in the cross-power spectrum are within the errors on data. In deeper surveys like H-ATLAS, the dependence of magnification bias on amplitude $A$ is more prominent (see fig.~18 of \citeauthor{Bianchini2015} \citeyear{Bianchini2015}).

\subsection{Median redshift}

If the errors on redshift are large, it can lead to systematic biases in measurements and estimation of parameters from datasets. The distribution of fractional error $\frac{\sigma_{z}}{1+z}$ is restricted to $\sim 0.15$ for NGP and HS-82, and $\sim 0.25$ for SGP Part-1 and Part-2. Such errors can shift the peak and median of the distribution significantly from its true position. In this section, we examine the robustness of our results quoted in Tab.~\ref{tab:likeli_result_original_kg} to the median redshift of the HELP catalogue which may be misestimated due to these systematic effects. For this test, it is convenient to use some model for the redshift distributions of the HELP catalogue and check the sensitivity of the cross-correlation amplitude and galaxy bias to the parameter related to median redshift. We model the redshift distributions by a function of the form
\begin{equation}
    \frac{dN}{dz} = a_{0}z^{a_{1}}\text{exp}\bigg[-\bigg(\frac{z}{a_{2}}\bigg)^{a_{3}}\bigg]
    \label{eq:redshift_model_function}
\end{equation}
We fit parameters $a_{0},a_{1},a_{2},a_{3}$ to the observed redshift distributions and their best fit values are given in Table \ref{tab:best_fit_values_redshift_modelling}.
\begin{table}
    \centering
    \caption{Best fit values of $a_{0},a_{1},a_{2},a_{3}$ for the modelling function given by Eq. \ref{eq:redshift_model_function}.}    
    \begin{tabular}{c|c|c|c|c}
        \hline\hline
        Patch & $a_{0}$ & $a_{1}$ & $a_{2}$ & $a_{3}$ \\
        \hline
        NGP & 5.843 & 1.007 & 0.602 & 3.014 \\
        HS-82 & 8.419 & 1.756 & 0.579 & 1.733 \\
        SGP Part-1 & 15.556 & 1.925 & 0.415 & 1.320 \\
        SGP Part-2 & 9.553 & 1.776 & 0.552 & 1.659 \\
        \hline
    \end{tabular}
    \label{tab:best_fit_values_redshift_modelling}
\end{table}

The parameter $a_{2}$ is different from the median redshift $z_{\rm median}$, but raising or lowering it will have a similar effect as shifting the median redshift of distribution. Thus, we use $a_{2}$ as a proxy for median redshift.

\begin{figure}
    \centering
    \includegraphics[width=\linewidth]{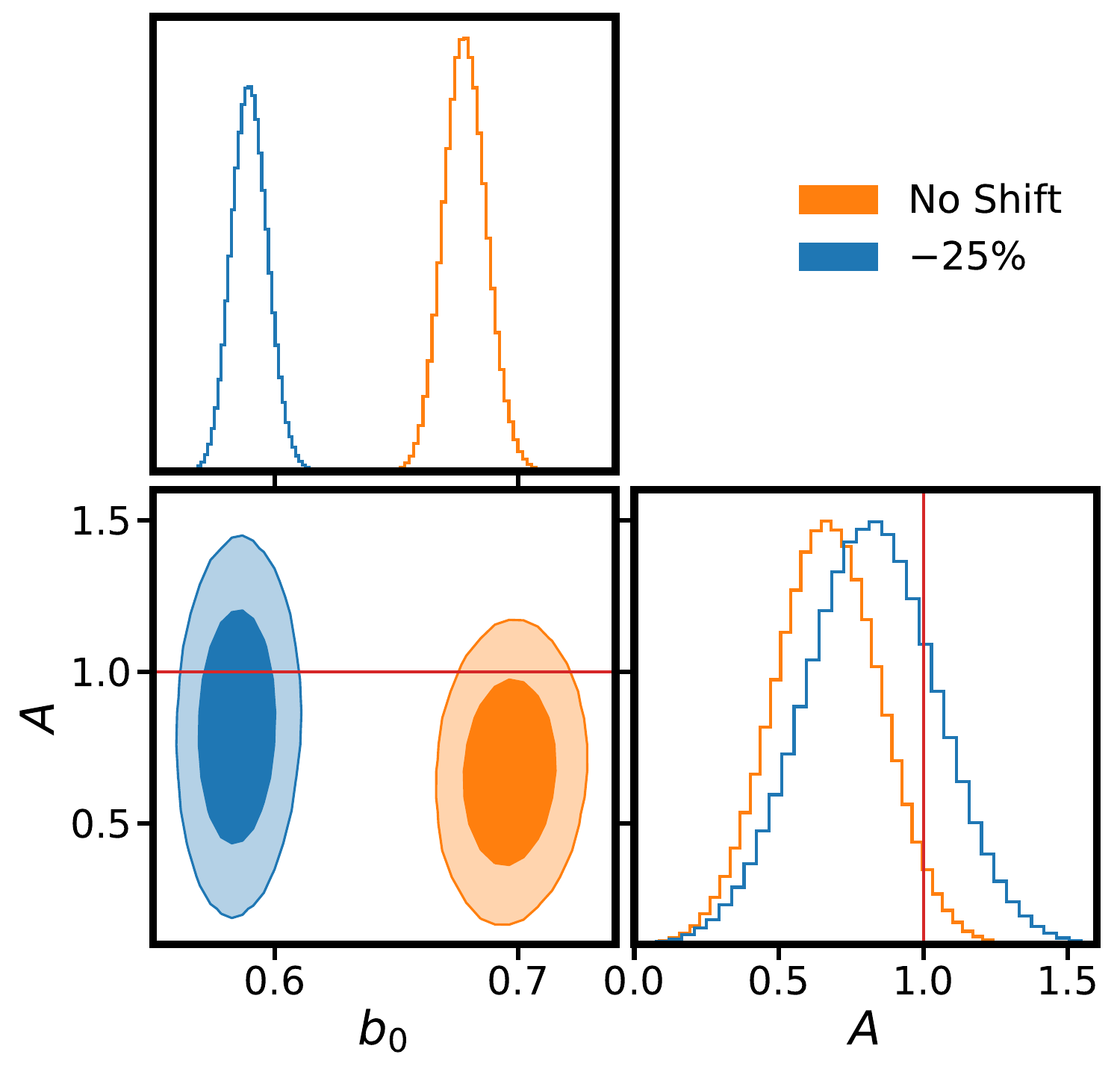}
    \caption{The parameters for SGP Part-2 computed with $-25\%$ shift in the median redshift of the distribution and without any shift. We show $68\%$ and $95\%$ contours with darker and lighter shaded regions, respectively. The red line represent the expected value $A=1$.}
    \label{fig:testing_depth_plot_sgp_part2}
\end{figure}

By changing the values of parameter $a_{2}$, we recompute the theoretical power spectrum and re-estimate the parameters $b_{0}$ and $A$. We find that for NGP and SGP Part-1 we need around $10\,\%$, HS-82 needs $\sim 20\,\%$ and for SGP Part-2 we need a median redshift around $25\,\%$ lower than that estimated to alleviate the tension on amplitude $A$. Fig.~\ref{fig:testing_depth_plot_sgp_part2} shows the comparison of the 1- and 2-dimensional posteriors for SGP Part-2 field computed with $25\,\%$ smaller value of the median redshift and without any change of the median.

To get an idea, of whether such significant changes of the median redshift are possible for catalogue with errors of the photometric redshift of the order of $\frac{\sigma_{z}}{1+z} \sim 0.2$, we create mock catalogues of objects assuming some true redshift distribution and then randomly redistribute redshift of each object according to Gaussian probability function with standard deviation corresponding to photometric redshift errors. We generate in this way a few hundred mock catalogues and estimate the average redshift distribution for them. In the end, we compare the median redshift of the true distribution and the average one. We find that the median of the average distribution, which takes into account photometric redshift errors, is $\sim 5-7\,\%$ larger than the median of the true distribution. This difference is too small to bring the amplitude measured for HELP patches consistent with unity.

\subsection{Photometric calibration errors}

Photometric calibration errors are systematics that can cause the magnitude limit of a survey to vary across the sky. This can cause changes in the number density coming from the survey across the sky. The variation in number density which does not correspond to fluctuations in physical matter density biases the galaxy over-density power spectrum and in turn may affect the estimation of parameters. The photometric calibration error is important at large scales, but since there are different surveys combined to make the HELP catalogue, we can expect significant effects at smaller angular scales also. The calibration error can, thus, be one of the major challenges in cross-correlation studies using catalogues combined from different surveys.

We follow the procedure outlined in \cite{Huterer2013} to study the effect of calibration errors using simulations. The true galaxy number counts in a given direction of sky $N_{\text{true}}(\hat{\textbf{n}})$ is affected by the calibration field $c(\hat{\textbf{n}})$
\begin{equation}
    N_{\text{obs}}(\hat{\textbf{n}}) = [1+c(\hat{\textbf{n}})]\,N_{\text{true}}(\hat{\textbf{n}})
    \label{eq:calib_error_true_obs_num_counts}
\end{equation}
The calibration field $c(\hat{\textbf{n}})$ can be computed using the relation
\begin{equation}
    c(\hat{\textbf{n}}) = \ln{(10)}\,s(z)\,\delta m(\hat{\textbf{n}})
\end{equation}
where $\delta m$ is the is variation in the magnitude limit in the direction $\hat{\textbf{n}}$ for some waveband and $s(z)$ is the faint end slope of the luminosity function given by
\begin{equation}
    s(z) \equiv \frac{\text{d}\log_{10}N(z,>m)}{\text{d}m}\Bigr|_{m_{\rm max}}
\end{equation}
where $m_{\rm max}$ is the maximal apparent magnitude for a given waveband.

\begin{figure*}
    \begin{subfigure}{.48\linewidth}
        \centering
        \includegraphics[width=\linewidth]{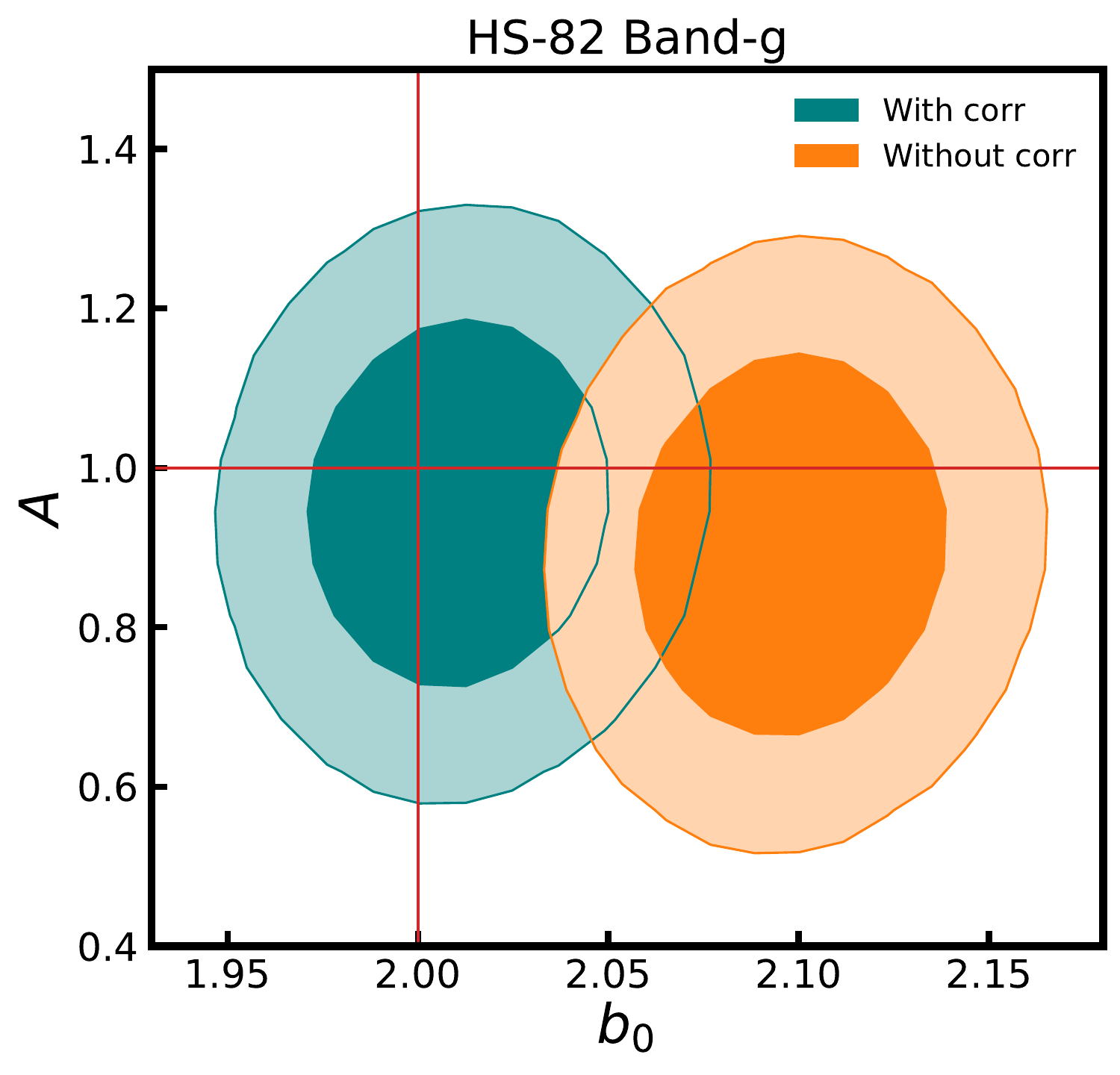}
        \captionsetup{labelformat=empty}
    \end{subfigure}%
    \begin{subfigure}{.48\linewidth}
        \centering
        \includegraphics[width=\linewidth]{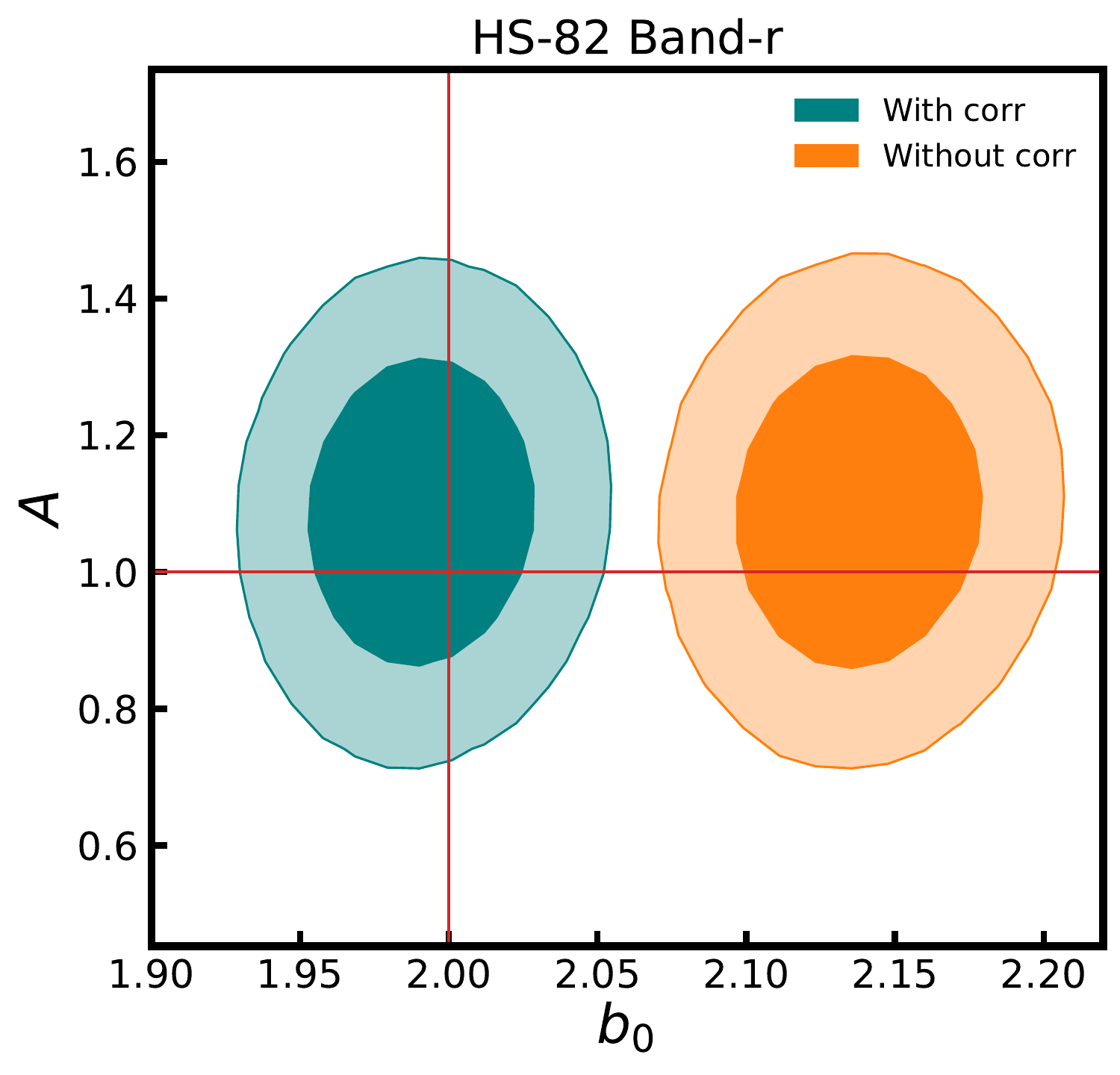}
        \captionsetup{labelformat=empty}
    \end{subfigure}
    \caption{Comparison of estimated parameters with and without correction for the calibration error for $g$ and $r$ bands using simulations corresponding to HS-82 field. We show $68\,\%$ and $95\,\%$ contours with darker and lighter shaded regions, respectively. The red lines represent the true values of $b_{0}$ and $A$ parameters used in simulations, i.e.~$b_{0}=2$ and $A=1$.}
    \label{fig:hs82_params_calib_error_band_g_and_band_r}
\end{figure*}

For HELP fields, we study the effects of calibration error for bands $g$ and $r$ as these bands are the deepest in magnitude and we also have maximum coverage for them. For a given field we simulate galaxy number count maps which include calibration field $c(\mathbf{\hat{n}})$ according to Eq.~\ref{eq:calib_error_true_obs_num_counts}. Then, we estimate parameters for these simulations without correction and after correcting for the calibration errors using Eq.~\ref{eq:calib_error_true_obs_num_counts}. The estimated parameters for the HS-82 field are shown in Fig.~\ref{fig:hs82_params_calib_error_band_g_and_band_r}. As we can notice the cross-correlation amplitude remains unchanged, while the galaxy linear bias parameter is shifted up by $1-2$ standard deviations. A similar effect is seen also for the remaining fields.

This test using simulations shows that amplitude values are robust with respect to photometric calibration errors. Nevertheless, even if they had significant impact, correction for the calibration errors would require quite a precise estimation of the calibration field. In the case of simulations, we assumed that the calibration field is known, while for real data we have only a tentative estimation of the field. For that reason, we do not apply this correction in the analysis of data.

\subsection{Catastrophic photo-$z$ error rate}

In photo-$z$ surveys, galaxies are often subject to catastrophic photo-$z$ errors where the true value of redshift is misestimated by a significant amount. The reasons for this are not fully understood, but, like the conventional photo-$z$ error case, the rate and outcome of catastrophic errors depend strongly on the number of photometric filters and their relation to the spectral features that carry principal information about the redshift \citep{Muir2016}.

\begin{figure}
    \centering
    \includegraphics[width=\linewidth]{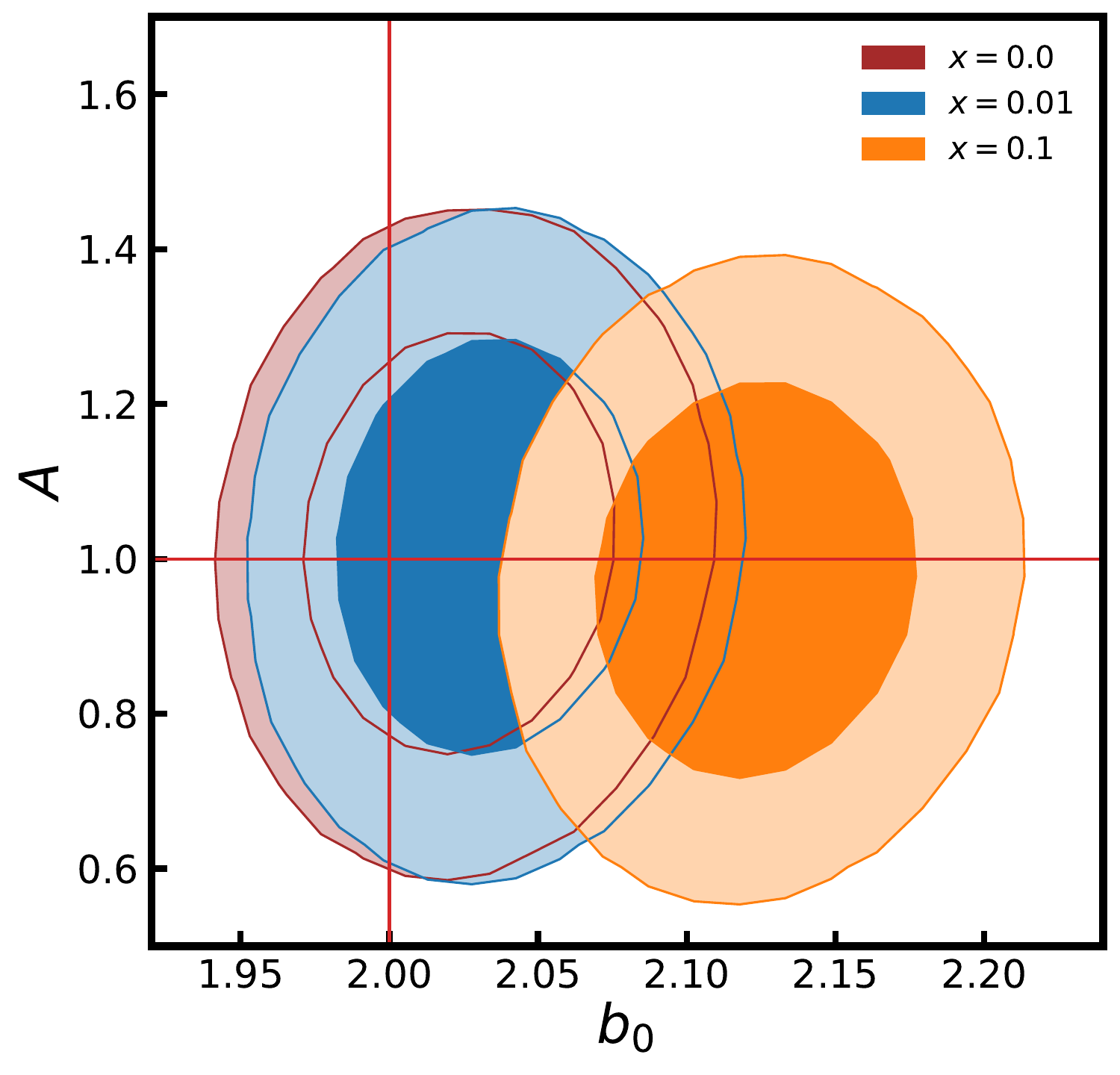}
    \caption{Effect of catastrophic errors on the inferred values of cross-correlation amplitude $A$ and galaxy linear bias parameter $b_{0}$ for SGP Part-2. We show $68\%$ and $95\%$ contours with darker and lighter shaded regions, respectively, for $x=0$ (no catastrophic errors), $x=0.01$ and $x=0.1$ catastrophic error rate. The red lines represent the true values of $b_{0}$ and $A$ used in simulations ($b_{0}=2$ and $A=1$).}
    \label{fig:testing_cat_err_plot_sgp_part2}
\end{figure}

We model catastrophic photo-$z$ errors by randomly assigning estimated redshifts for a fraction $x$ of galaxies in the HELP catalogue. For tests, we chose two different fractions of catastrophic photo-$z$ error rates, $x=0.01 \text{ and } 0.1$ which roughly amounts to the lower and upper limits, respectively, of the fraction achieved in current surveys (see \citeauthor{Xiao2022} \citeyear{Xiao2022}; \citeauthor{Jouvel2017} \citeyear{Jouvel2017}; \citeauthor{Muir2016} \citeyear{Muir2016}). We take the range of randomized redshifts to be $z \in [0.01,3.0]$. With randomly assigned estimated redshifts for fraction $x$ of galaxies, we compute the redshift distribution and, hence, simulate galaxy over-density maps. We then estimate parameters $b_{0}$ and $A$ from these simulated maps using theoretical power spectrum computed by assuming redshift distribution without catastrophic errors.  Fig. \ref{fig:testing_cat_err_plot_sgp_part2} shows contour plots in $(b_{0},A)$ plane for simulated SGP Part-2 field corresponding to $x=0.01 \text{ and } 0.1$. We notice that different fractions of catastrophic photo-$z$ errors have no notable effect on the amplitude of cross-correlation $A$ for the HELP catalogue. We see a similar trend also for other patches.

\section{Summary}\label{sec:summary}

We have presented measurement of cross-correlation between the minimum-variance and SZ-deproj CMB lensing convergence map from \textit{Planck} 2018 data release and galaxy catalogues from the \textit{Herschel} Extragalactic Legacy Project. For our analysis, we have selected three of the largest and most uniform fields of the catalogue namely, NGP, HS-82, and SGP divided into two parts. The areas covered by these fields are: $\sim 180$ deg$^2$, $\sim 255$ deg$^2$, $\sim 85$ deg$^2$ and $\sim 145$ deg$^2$, respectively.

We have shown that for MV lensing map the no correlation hypothesis can be ruled out with a significance of about $1.7\,\sigma$ for NGP, $\sim 9.2\,\sigma$ for HS-82, $\sim 2.6\,\sigma$ for SGP Part-1 and $\sim 2.5\,\sigma$ for SGP Part-2 field. A joint analysis of galaxy auto-power spectrum and cross-power spectrum using Maximum Likelihood approach gives the galaxy linear bias parameter for different fields ranging from $b_{0}=0.70 \pm 0.01$ for SGP Part-2 to $b_{0}=1.02 \pm 0.02$ for SGP Part-1 field. The cross-correlation amplitude varies from $A=0.67 \pm 0.18$ for SGP Part-2 to $A=0.80 \pm 0.23$ for SGP Part-1 field and a significance of its deviation from one is ~1\,$\sigma$ for NGP and SGP Part-1, $\sim 1.5\,\sigma$ for HS-82 and $\sim 2\,\sigma$ for SGP Part-2. For the SZ-deproj CMB lensing map the amplitude is higher and consistent with one for all fields, except NGP field for which its value is lower than for the MV lensing map.

Though, a significance of the deviation for the MV lensing map is not very high, especially for NGP and SGP Part-1 fields, in all cases the amplitude is biased towards lower values suggesting that there is some systematic error in the analysis.
To check it we have investigated some systematic errors that can account for this deviation, such as the effect of magnification bias caused by weak gravitational lensing and catastrophic photo-$z$ errors which were found to have no notable improvement over the detected tension of cross-correlation amplitude $A$. We also examined the effect of shifting the estimated median redshift of HELP galaxies and concluded that a lower effective median redshift can increase the estimated value of amplitude $A$, suggesting that the HELP catalogue may be shallower than expected. However, the amount of shift required to remove the observed tension on the amplitude of cross-correlation, i.e.~20-25\,\% of the median redshift, is much larger than the potential offset related with photometric redshift errors, i.e.~5\,\% of the median. We also found out that variations across the fields in magnitude limits of the catalogue caused by photometric calibration errors has no significant effect on the cross-correlation amplitude.

To conclude, the amplitude turned out to be robust with respect to all studied systematic errors. Only in the case of using SZ-deproj lensing map do we observe stronger cross-correlations and higher amplitude which is then consistent with one for all fields except the NGP field, which shows a weaker correlation than for MV map. These disparities can be explained by differences between MV and SZ-deproj lensing maps, however, it also shows that we need a more robust estimation of the CMB lensing map for cross-correlation studies. We can expect that forthcoming CMB experiments and galaxy surveys will allow us to perform more robust and precise cross-correlation measurements in the future.

\section*{Acknowledgements}
The authors would like to thank Raphael Shirley, Kenneth Duncan, and Katarzyna Ma\l{}ek for their help with different aspects of HELP data. We thank Federico Bianchini for his suggestion on the jackknifing approach to the galaxy shot noise and valuable comments. We also thank Agnieszka Pollo for valuable comments and discussions. We thank the anonymous reviewer for careful reading of the manuscript and their helpful and relevant comments which allowed us to significantly improve this paper. The work has been supported by the Polish Ministry of Science and Higher Education grant DIR/WK/2018/12 and is based on observations obtained with Planck (http://www.esa.int/Planck), an ESA science mission with instruments and contributions directly funded by ESA Member States, NASA, and Canada. The authors acknowledge the use of CAMB and HEALPix packages.

\section*{Data Availability}
The HELP catalogue used in our analysis is publicly available at \url{http://hedam.lam.fr/HELP/dataproducts/} and the \textit{Planck} 2018 data products can be obtained from \url{https://pla.esac.esa.int/\#cosmology}. The simulated data sets will be shared on reasonable request to the corresponding author.




\bibliographystyle{mnras}
\bibliography{cross-correlation} 




\appendix
\onecolumn

\section{Covariance Matrix}\label{apndx:covariance_matrix}

We discuss here in detail the expression of covariance presented in Eq. \ref{eq:error_covariance}. We start from the pseudo covariance given by

\begin{equation}
    \begin{split}
    \widetilde{Cov}_{\ell\ell '}^{AB,CD}  &=  \langle (\langle\tilde{C}_{\ell}^{AB}\rangle - \tilde{C}_{\ell}^{AB})(\langle\tilde{C}_{\ell '}^{CD}\rangle - \tilde{C}_{\ell '}^{CD})\rangle\\
    &= \langle \tilde{C}_{\ell}^{AB}\tilde{C}_{\ell '}^{CD} \rangle - \langle \tilde{C}_{\ell}^{AB}\rangle\langle\tilde{C}_{\ell '}^{CD} \rangle
    \end{split}
\label{eq_apndx:pseudo_cl}
\end{equation}
where $\tilde{C}_{\ell}$ is pseudo power spectrum and $A,B,C,D$ represent scalar fields on sky. Let $\tilde{a}_{\ell m}$ be the spherical harmonic coefficients of $\tilde{C}_{\ell}$.

\begin{equation}
\begin{split}
    \widetilde{Cov}_{\ell\ell '}^{AB,CD}  &= \frac{1}{(2\ell+1)(2\ell'+1)} \sum_{mm'}\bigg[\langle \tilde{a}_{\ell m}^{A}\tilde{a}_{\ell m}^{B*}\tilde{a}_{\ell' m'}^{C}\tilde{a}_{\ell' m'}^{D*} \rangle - \langle \tilde{a}_{\ell m}^{A}\tilde{a}_{\ell m}^{B*}\rangle\langle\tilde{a}_{\ell' m'}^{C}\tilde{a}_{\ell' m'}^{D*} \rangle\bigg]\\
    &= \frac{1}{(2\ell+1)(2\ell'+1)} \sum_{mm'}\bigg[\langle \tilde{a}_{\ell m}^{A}\tilde{a}_{\ell' m'}^{C*}\rangle\langle\tilde{a}_{\ell' m'}^{D}\tilde{a}_{\ell m}^{B*} \rangle + \langle \tilde{a}_{\ell m}^{A}\tilde{a}_{\ell' m'}^{D*}\rangle\langle\tilde{a}_{\ell' m'}^{C}\tilde{a}_{\ell m}^{B*} \rangle\bigg]
\end{split}
\label{eq_apndx:pseudo_alm}
\end{equation}
We can express $\tilde{a}_{\ell m}$ in terms of $a_{\ell m}$, the spherical harmonic coefficients of full sky power spectrum $C_{\ell}$, using the mode-mode coupling kernel $K_{\ell m\ell ' m'}$ \citep{Hivon2002} as:

\begin{equation}
    \tilde{a}_{\ell m} = \sum_{\ell 'm'}a_{\ell' m'}K_{\ell m\ell ' m'}
\end{equation}
with which Eq. \ref{eq_apndx:pseudo_alm} becomes

\begin{equation}
\begin{split}
    \widetilde{Cov}_{\ell\ell '}^{AB,CD}  = \frac{1}{(2\ell+1)(2\ell'+1)}\sum_{mm'} \sum_{\substack{\ell_{1}\ell_{2}\ell_{3}\ell_{4}\\m_{1}m_{2}m_{3}m_{4}}}
        \bigg[&\langle a_{\ell_{1} m_{1}}^{A} a_{\ell_{2} m_{2}}^{C*}\rangle K_{\ell m\ell_{1}m_{1}}^{A}K_{\ell' m'\ell_{2}m_{2}}^{C*} \langle a_{\ell_{3} m_{3}}^{D} a_{\ell_{4} m_{4}}^{B*} \rangle K_{\ell' m'\ell_{3}m_{3}}^{D}K_{\ell m\ell_{4}m_{4}}^{B*}\\
        &+\langle a_{\ell_{1} m_{1}}^{A} a_{\ell_{3} m_{3}}^{D*}\rangle K_{\ell m\ell_{1}m_{1}}^{A}K_{\ell' m'\ell_{3}m_{3}}^{D*} \langle a_{\ell_{2}m_{2}}^{C}a_{\ell_{4} m_{4}}^{B*} \rangle K_{\ell' m'\ell_{2}m_{2}}^{C}K_{\ell m\ell_{4}m_{4}}^{B*}\bigg]
        \label{eq_apndx:pseudo_full_alm}
\end{split}
\end{equation}
Using $\langle a_{\ell m}a_{\ell' m'} \rangle = \delta_{\ell\ell'}\delta_{mm'}\langle C_{\ell}\rangle$, in Eq. \ref{eq_apndx:pseudo_full_alm}, we get

\begin{equation}
    \widetilde{Cov}_{\ell\ell '}^{AB,CD} = \frac{1}{(2\ell+1)(2\ell'+1)}\sum_{mm'} \sum_{\substack{\ell_{1}\ell_{4}\\m_{1}m_{4}}}
        \bigg[\langle C_{\ell_{1}}^{AC}\rangle\langle C_{\ell_{4}}^{DB}\rangle K_{\ell m \ell_{1} m_{1}}^{A} K_{\ell' m'\ell_{1}m_{1}}^{C*} K_{\ell' m'\ell_{4}m_{4}}^{D} K_{\ell m \ell_{4} m_{4}}^{B*} + \langle C_{\ell_{1}}^{AD}\rangle\langle C_{\ell_{4}}^{CB}\rangle K_{\ell m \ell_{1} m_{1}}^{A} K_{\ell' m'\ell_{1}m_{1}}^{D*} K_{\ell' m'\ell_{4}m_{4}}^{C} K_{\ell m \ell_{4} m_{4}}^{B*}\bigg]
        \label{eq_apndx:pseudo_cov_without_fullsky_approx}
\end{equation}
We develop each term in Eq. \ref{eq_apndx:pseudo_cov_without_fullsky_approx} assuming the large sky coverage \citep{Efstathiou2004}:

\begin{equation}
    \sum_{\substack{\ell_{1}\ell_{4}\\m_{1}m_{4}}} \langle C_{\ell_{1}}^{AC}\rangle\langle C_{\ell_{4}}^{DB}\rangle K_{\ell m \ell_{1} m_{1}}^{A} K_{\ell' m'\ell_{1}m_{1}}^{C*} K_{\ell' m'\ell_{4}m_{4}}^{D} K_{\ell m \ell_{4} m_{4}}^{B*} = \sqrt{C_{\ell}^{AC}C_{\ell'}^{AC}C_{\ell}^{DB}C_{\ell'}^{DB}}\sum_{\substack{\ell_{1}\ell_{4}\\m_{1}m_{4}}}K_{\ell m \ell_{1} m_{1}}^{A} K_{\ell' m'\ell_{1}m_{1}}^{C*} K_{\ell' m'\ell_{4}m_{4}}^{D} K_{\ell m \ell_{4} m_{4}}^{B*}
    \label{eq_apndx:pseudo_cov_with_fullsky_approx}
\end{equation}
Expanding the mode-mode coupling kernels in terms of sum over pixels and then, applying the completeness relation of spherical harmonics:

\begin{equation}
\begin{split}
    \sum_{\ell_{1}m_{1}}K_{\ell m \ell_{1} m_{1}}^{X} K_{\ell' m'\ell_{1}m_{1}}^{Y*} &=  \sum_{\ell_{1}m_{1}} \sum_{pq} w_{p}^{X}w_{q}^{Y*}\Omega_{p}\Omega_{q} Y_{\ell m}(\theta_{p}) Y_{\ell_{1} m_{1}}^{*}(\theta_{p}) Y_{\ell_{1} m_{1}}(\theta_{q}) Y_{\ell' m'}^{*}(\theta_{q})\\
    &= \sum_{pq}w_{p}^{X}w_{q}^{Y*}\Omega_{p}\Omega_{q} Y_{\ell m}(\theta_{p})Y_{\ell' m'}^{*} (\theta_{q})\frac{\delta(\theta_{p}-\theta_{q})}{\Omega_{q}}\\
    &= \sum_{p}w_{p}^{XY}\Omega_{p} Y_{\ell m}(\theta_{p})Y_{\ell' m'}^{*}(\theta_{p})\\
    &= K_{\ell m\ell'm'}^{XY}
\end{split}
\label{eq_apndx:composite_kernel}
\end{equation}
where $w_{p}$ is an arbitrary weight function, $\Omega_{p}$ is area of each pixel and we have defined $w_{p}^{XY} = w_{p}^{X}w_{p}^{Y*}$ as another arbitrary weight function. With Eq. \ref{eq_apndx:pseudo_cov_with_fullsky_approx} and Eq. \ref{eq_apndx:composite_kernel}, Eq. \ref{eq_apndx:pseudo_cov_without_fullsky_approx} simplifies as:

\begin{equation}
    \widetilde{Cov}_{\ell\ell '}^{AB,CD} = \frac{1}{(2\ell+1)(2\ell'+1)}\sum_{mm'} \bigg[\sqrt{C_{\ell}^{AC}C_{\ell'}^{AC}C_{\ell}^{DB}C_{\ell'}^{DB}}K_{\ell m\ell'm'}^{AC}K_{\ell m\ell'm'}^{BD*}+\sqrt{C_{\ell}^{AD}C_{\ell'}^{AD}C_{\ell}^{CB}C_{\ell'}^{CB}}K_{\ell m\ell'm'}^{AD}K_{\ell m\ell'm'}^{BC*}\bigg]
    \label{eq_apndx:pseudo_covariance}
\end{equation}
The product of coupling kernels can be expanded in terms of Wigner-3j symbols as:

\begin{equation}
\begin{split}
    \sum_{mm'} K_{\ell m\ell'm'}^{AC}K_{\ell m\ell'm'}^{BD*} =& \sum_{mm'}\sum_{\substack{\ell_{1}\ell_{2}\\m_{1}m_{2}}}w_{\ell_{1}m_{1}}^{AC}w_{\ell_{2}m_{2}}^{BD*}
    \frac{(2\ell+1)(2\ell'+1)}{4\pi}\sqrt{(2\ell_{1}+1)(2\ell_{2}+1)}\\
    &\times\begin{pmatrix} \ell &\ell' &\ell_{1}\\ 0 &0 &0\end{pmatrix}\begin{pmatrix} \ell &\ell' &\ell_{2}\\ 0 &0 &0\end{pmatrix}\begin{pmatrix} \ell &\ell' &\ell_{1}\\ m &-m' &m_{1}\end{pmatrix} \begin{pmatrix} \ell &\ell' &\ell_{2}\\ m &-m' &m_{2}\end{pmatrix}
\end{split}
\label{eq_apndx:coupling_matrix_wigner_expansion}
\end{equation}
Using the orthogonality relations of Wigner-3j symbols, Eq. \ref{eq_apndx:coupling_matrix_wigner_expansion} simplifies as

\begin{equation}
    \sum_{mm'} K_{\ell m\ell'm'}^{AC}K_{\ell m\ell'm'}^{BD*} = (2\ell+1)M_{\ell\ell'}^{AC,BD}
\end{equation}
where $M_{\ell\ell'}^{AB,CD}$ is given by \citep{Hivon2002}
\begin{equation}
    M_{\ell\ell'}^{AB,CD} = \frac{2\ell'+1}{4\pi}\sum\limits_{\ell_{1}}(2\ell_{1}+1)\bigg[\frac{1}{2\ell_{1}+1}\sum\limits_{m_{1}}w_{\ell_{1}m_{1}}^{AC}w_{\ell_{2}m_{2}}^{BD*}\bigg]\begin{pmatrix} \ell &\ell' &\ell_{1}\\ 0 &0 &0\end{pmatrix}^{2}
\end{equation}
This transforms the expression for pseudo covariance matrix Eq. \ref{eq_apndx:pseudo_covariance} as

\begin{equation}
    \widetilde{Cov}_{\ell\ell '}^{AB,CD} = \frac{1}{(2\ell'+1)} \bigg[M_{\ell\ell'}^{AC,BD} \sqrt{C_{\ell}^{AC}C_{\ell'}^{AC}C_{\ell}^{DB}C_{\ell'}^{DB}} + M_{\ell\ell'}^{AD,BC} \sqrt{C_{\ell}^{AD}C_{\ell'}^{AD}C_{\ell}^{CB}C_{\ell'}^{CB}}\bigg]
\end{equation}
The binned covariance matrix for full-sky is given by \citep{Brown2005}:

\begin{equation}
    Cov_{LL'}^{AB,CD} = (M_{LL_{1}}^{AB^{-1}}P_{L_{1}\ell})\bigg[\frac{M_{\ell\ell'}^{AC,BD} \sqrt{C_{\ell}^{AC}C_{\ell'}^{AC}C_{\ell}^{DB}C_{\ell'}^{DB}} + M_{\ell\ell'}^{AD,BC} \sqrt{C_{\ell}^{AD}C_{\ell'}^{AD}C_{\ell}^{CB}C_{\ell'}^{CB}}}{(2\ell'+1)} \bigg](M_{L'L_{2}}^{CD^{-1}}P_{L_{2}\ell'})^{T}
    \label{eq_apndx:full_cov}
\end{equation}
Eq. \ref{eq_apndx:full_cov} is similar to that obtained by \cite{Tristram2005}. This expression takes into account different fractions of sky covered by the fields $A,B,C,D$.

\section{HELP survey details}

\begin{table}
	\centering
	\captionsetup{justification=centering}
	\caption{Fraction of objects for a given HELP field observed by different surveys and photometric filters.}
	\label{apndx_tab:bandwise_coverage_help_fields}
	\begin{tabular}{lccccc} 
		\hline\hline
		Survey & Filter & \multicolumn{4}{c}{$\%$ of objects}\\
		\cline{3-6}
		 & & NGP & HS-82 & SGP Part-1 & SGP Part-2\\
		\hline\noalign{\vskip 0.2cm}
		\multirow{3}{4em}{DECaLS} & g & 15.9 & - & - & -\\
		& r & 19.4 & - & - & -\\
		& z & 11.6 & - & - & -\\
		\hline\noalign{\vskip 0.2cm}
		\multirow{5}{4em}{DES} & g & - & 99.3 & 2.0 & 98.8\\
		& r & - & 99.5 & 2.2 & 99.8\\
		& i & - & 98.7 & 2.2 & 99.7\\
		& z & 86.1 & 99.3 & 2.0 & 99.3\\
		& y & - & 95.1 & 2.0 & 92.8\\
		\hline\noalign{\vskip 0.2cm}
		\multirow{4}{4em}{KiDS} & u & - & - & 90.2 & 1.2\\
		& g & - & - & 98.7 & 13.0\\
		& r & - & - & 99.8 & 19.6\\
		& i & - & - & 99.3 & 17.6\\
		& z & - & - & 16.6 & 7.7\\
		\hline\noalign{\vskip 0.2cm}
		\multirow{5}{6em}{PanSTARRS} & g & 99.3 & 43.8 & 0 & 9.4\\
		& r & 100 & 45.7 & 0 & 10.2\\
		& i & 100 & 46.0 & 0 & 10.6\\
		& z & 100 & 45.7 & 0 & 10.2\\
		& y & 99.6 & 44.2 & 0 & 9.6\\
		\hline\noalign{\vskip 0.2cm}
		\multirow{5}{6em}{RCSLenS} & g & - & 20.6 & - & -\\
		& r & - & 23.9 & - & -\\
		& i & - & 3.8 & - & -\\
		& z & - & 20.9 & - & -\\
		& y & - & 9.3 & - & -\\
		\hline\noalign{\vskip 0.2cm}
		\multirow{5}{6em}{SDSS} & u & - & 35.8 & - & -\\
		& g & - & 35.8 & - & -\\
		& r & - & 35.8 & - & -\\
		& i & - & 35.8 & - & -\\
		& z & - & 35.8 & - & -\\
		\hline\noalign{\vskip 0.2cm}
		SHELA + SpIES & IRAC12 & - & 26.1 & - & -\\
		\hline\noalign{\vskip 0.2cm}
		\multirow{4}{6em}{UKIDSS-LAS} & Y & 59.5 & 10.6 & - & -\\
		& J & 50.0 & 8.1 & - & -\\
		& H & 56.8 & 10.0 & - & -\\
		& K & 58.9 & 13.7 & - & -\\
		\hline\noalign{\vskip 0.2cm}
		\multirow{5}{6em}{VISTA} & Y & - & 2.9 & 29.9 & 20.8\\
		& J & - & 23.0 & 34.5 & 23.0\\
		& H & - & 13.9 & 29.2 & 16.8\\
		& Ks & - & 22.7 & 31.1 & 18.3\\
		& Z & - & - & 51.3 & 34.9\\
		\hline  
	\end{tabular}
\end{table}


\bsp	
\label{lastpage}

\end{document}